\documentclass[review]{elsarticle}

\usepackage{lineno,hyperref}
\modulolinenumbers[5]
\usepackage{xcolor}
\usepackage[section]{placeins}
\usepackage{multirow}
\usepackage{url}
\usepackage{acronym}
\usepackage{amsmath}
\usepackage{amssymb}
\usepackage{graphicx}
\usepackage{caption}
\usepackage{mathtools, nccmath}
\usepackage{marginnote}
\usepackage{arydshln}
\usepackage{rotating}
\usepackage{natbib}
\usepackage{longtable}
\journal{Journal of the Association for Information Science and Technology}

\begin{document}

\begin{frontmatter}

\title{Learning to Rank from Relevance Judgments Distributions}

\author{Alberto Purpura \corref{mycorrespondingauthor}}
\cortext[mycorrespondingauthor]{Corresponding author.}
\ead{purpuraa@dei.unipd.it}

\author{Gianmaria Silvello}
\ead{gianmaria.silvello@unipd.it}

\author{Gian Antonio Susto}
\ead{gianantonio.susto@unipd.it}

\address{Department of Information Engineering, University of Padua, Italy.}

\begin{abstract}
\ac{LETOR} algorithms are usually trained on annotated corpora where a single relevance label is assigned to each available document-topic pair. Within the Cranfield framework, relevance labels result from merging either multiple expertly curated or crowdsourced human assessments. In this paper, we explore how to train \ac{LETOR} models with relevance judgments distributions (either real or synthetically generated) assigned to document-topic pairs instead of single-valued relevance labels. We propose five new probabilistic loss functions to deal with the higher expressive power provided by relevance judgments distributions and show how they can be applied both to neural and \acf{GBM} architectures.
Moreover, we show how training a \ac{LETOR} model on a sampled version of the relevance judgments from certain probability distributions can improve its performance when relying either on traditional or probabilistic loss functions. Finally, we validate our hypothesis on real-world crowdsourced relevance judgments distributions. 
Overall, we observe that relying on relevance judgments distributions to train different LETOR models can boost their performance and even outperform strong baselines such as LambdaMART on several test collections.
\end{abstract}

\begin{keyword}
information retrieval\sep learning to rank \sep machine learning
\end{keyword}

\end{frontmatter}


\section{Introduction}\label{sec:introduction}

\textit{\textbf{Motivation.}} Ranking is a problem that we encounter in a number of tasks we perform every day: from searching on the Web to online shopping. 
Given an unordered set of items, this problem consists of ordering the items according to a certain notion of relevance. Generally, in \ac{IR} we rely on a notion of relevance that depends on the information need of a user, expressed through a keyword query. 
When creating a new experimental collection, the corresponding relevance judgments are obtained by asking different judges to assign a relevance score to each document-topic pair. Multiple judges -- either trained experts or participants of a crowdsourcing experiment -- usually assess the same document-topic pair, and the final relevance label for the pair is obtained by aggregating these scores~\cite{hosseini2012aggregating}. This process is a cornerstone for system training and evaluation and has contributed to the continuous development of \ac{IR}, especially in the context of international evaluation campaigns. Nonetheless, the opinion of different judges on the same document-topic pair might be very different or even diverge to the opposite ends of the spectrum -- either because of random human errors or due to a different interpretation of a topic. Inevitably, the aggregation process conflates the multiple assessors viewpoints on document-topic pairs onto a single one, thus losing some information -- even though it also reduces annotation errors and outliers. 
Our research hypothesis is that \ac{ML} models -- i.e., \ac{LETOR}~\cite{tax2015cross} and \ac{NIR}~\cite{onal2018neural} models -- could use all the labels collected in the annotation process to improve the quality of their rankings. Indeed, judges disagreement on a certain document-topic pair can be due to an inherent difficulty of the topic or to the existence of multiple interpretations of it. We argue that designing \ac{ML} models able to learn from the whole distributions of relevance judgments could improve the models' representation of relevance and their performance through the usage of this additional information. 

\textit{\textbf{Methods.}} Following this idea, we propose to interpret the output of a \ac{LETOR} model as a probability value or distribution -- according to the experimental hypotheses -- and define different \ac{KL} divergence-based loss functions to train a model using a distribution of relevance judgments associated to the current training item. Such a training strategy allows us to leverage all the available information from human judges without additional computational costs compared to traditional \ac{LETOR} training paradigms. 

The loss functions we propose can be used to train any ranking model that relies on gradient-based learning, including popular \ac{NIR} models or \ac{LETOR} ones. 

{
In this work we focus on transformer-based neural \ac{LETOR} models and on one decision tree-based \ac{GBM} model -- the model at the base of the popular LambdaMART~\cite{burges2010ranknet} ranker and used as a strong baseline in many recent \ac{LETOR} research papers such as~\cite{bruch2019revisiting, bruch2019alternative, pasumarthi2020permutation, bruch2020stochastic, qin2021neural}.}

{
\textit{\textbf{Evaluation.}} We assess the quality of the proposed training strategies on four standard \ac{LETOR} collections (MQ2007, MQ2008, MSLR-WEB30K~\cite{qin2013introducing} and OHSUMED~\cite{qin2010letor}) and three different transformer-based \ac{LETOR} models. We also conduct a crowdsourcing experiment to build a new LETOR collection based on the COVID-19 MultiLingual Information Access (MLIA) data.~\footnote{\url{http://eval.covid19-mlia.eu}} We then use the raw relevance labels and their aggregated version to assess the impact on a LETOR model trained on raw relevance labels versus their aggregated form.}

\textit{\textbf{Contributions}} of this paper are (i) the definition of five new loss functions to train different LETOR models using probability distributions of relevance labels; (ii) an extensive evaluation of the loss functions on neural and decision tree-based \ac{LETOR} models using standard test collections and a newly created crowdsourced \ac{LETOR} test collection based on COVID-19 MLIA data.

\textit{\textbf{Outline.}} In Section \ref{sec:related}, we describe the most relevant training strategies for \ac{LETOR} and \ac{NIR} models; in Section \ref{sec:proposed}, we present the probabilistic loss functions  leveraging on relevance judgments distributions and the novel neural \ac{LETOR} model we employ for evaluation; in Section \ref{sec:setup}, we describe the experimental setup and in Section \ref{sec:eval} we discuss the evaluation results; in Section \ref{sec:conclusion}, we draw some conclusions and discuss future work.
\section{Related Work} \label{sec:related}
Decision tree-based approaches such as LambdaMART~\cite{burges2010ranknet} have been for many years the most popular \ac{ML} models in this domain, but recently -- thanks to the growing size of \ac{LETOR} collections, new optimization functions~\cite{bruch2020stochastic} and feature normalization strategies~\cite{zhuang2020feature} -- deep learning approaches, transformer-based ones in particular~\cite{qin2021neural}, are showing an increasingly competitive performance. 

{
Transformer-based~\ac{LETOR} models rely on one or more self-attention layers. This type of neural layer architecture was originally proposed in \cite{vaswani2017attention}, and later popularized by language models such as \ac{BERT}~\cite{devlin2018bert}. This architecture allows \ac{LETOR} models to efficiently evaluate and compare lists of candidate relevant documents to a user query, providing a numerical estimate of their relevance. One of the latest and most successful approaches of this kind is \ac{DASALC}~\cite{qin2021neural}. It relies on a few strategies such as neural feature transformation, self-attention layers, a listwise ranking loss and model ensembling, to outperform strong non-neural baselines such as LambdaMART on public \ac{LETOR} collections. Strategies such as neural feature transformation are frequently employed in the context of neural \ac{LETOR} models to normalize the representation of their inputs so that they could be better interpreted by such models~\cite{zhuang2020feature}. Self-attention layers -- such as the one we employ in the proposed neural \ac{LETOR} model described in section \ref{sec:proposed} -- allow to efficiently compare groups of items at the feature level. Finally, listwise ranking losses and model ensembling strategies are already popular solutions in \ac{LETOR} and machine learning to improve the performance of ranking models~\cite{burges2010ranknet, bruch2019revisiting}.\\
In this work, we will show how we can achieve similar improvements in a more efficient way through the usage of a new class of \textit{probabilistic} loss functions that we propose.\\
Indeed, in addition to the development of new architectures, another important branch of research in the \ac{LETOR} domain focuses on the study of new loss functions specific to ranking problems~\cite{tax2015cross}.}
These loss functions are generally categorized as \emph{pointwise}, \emph{pairwise} and \emph{listwise}~\cite{chen2009ranking}.
Pointwise loss functions are used to train a model to fit the corresponding relevance score for each document-topic pair as in a regression task. Loss functions belonging to this class can be described using the following general formulation: $\text{Pointwise}(q, d, y) = f(s(q, d), y)$,
where $q$ indicates a query, $d$ a document, $y$ its relevance label, $s(\cdot, \cdot)$ the function learned by a \ac{ML} model to compute the relevance of a document given a query, and $f(\cdot)$ the generic function which compares the score computed by the model that is being trained with the corresponding relevance label. One of the possible implementations of $f(\cdot)$ is the \ac{MSE}~\cite{liu2018leveraging}.
Pairwise loss functions consider pairs of documents and compare their relevance scores using different strategies. This class of losses can be formalized as: $\text{Pairwise}(q, d_1, d_2) = f(s(q, d_1), s(q, d_2)),$
where the function $f$ can have different formulations such as the Hinge function $\phi(z_1, z_2) = max(0, 1-z_1 + z_2)$, where $z_i = s(q, d_i)~\forall i \in \{1,..., n\}$~\cite{guo2016deep}.
Finally, listwise loss functions take into account a set of documents relative to a certain query and compute the loss for the group of items as: $\text{Listwise}(q, \{d_1, ..., d_n\}, \{y_1, ..., y_n\}) = f((s(q, d_1), ..., s(q, \\ d_n)), \{y_1, ..., y_n\}),$
where $\{y_1, ..., y_n\}$ are the relevance judgments associated to $\{d_1, ..., d_n\}$ and the function $f(\cdot)$ can take different formulations such as the
ApproxNDCG~\cite{qin2010general} loss.

Amongst the numerous loss function formulations proposed by the \ac{LETOR} community, the most widely-employed in the latest state-of-the-art text-based \ac{NIR}~\cite{macavaney2019cedr} and \ac{LETOR} models~\cite{zhuang2020feature} are the pairwise Hinge~\cite{onal2018neural} and the listwise ApproxNDCG loss~\cite{qin2010general, bruch2019revisiting}, respectively. 
The formulation of the Hinge loss as used in the \ac{NIR} domain is $\text{Hinge}(q, d^+, d^-) = \max (0, 1 - s(q, d^+) + s(q, d^-)),$
where $d^+$ and $d^-$ identify respectively a relevant and a not-relevant document for the query $q$. The goal of this loss function is to maximize the difference between the relevance probabilities -- indicated by the function $s(q, \cdot)$ -- computed for each document. The Hinge loss is frequently chosen to train text-based \ac{NIR} models such as DRMM~\cite{guo2016deep}, MatchPyramid~\cite{pang2016study} or the most recently proposed CEDR~\cite{macavaney2019cedr}, but was also used as a baseline in \ac{LETOR} research works such as in \cite{zamani2018theory}, where it demonstrated to be a competitive candidate among other pointwise or pairwise loss functions. The main advantage of this loss function is its ability to perform well on relatively small datasets -- as it is often the case for text-based \ac{NIR} models when evaluated on shared \ac{IR} test collections.
Differently from the Hinge loss, the ApproxNDCG loss can take into account more than two documents at a time and, as the name suggests, provides a differentiable approximation of the \ac{nDCG} measure for the evaluation of a ranked list. Given a permutation $\pi$ of items $\{x_1, ..., x_n\}$ and the corresponding sequence of relevance labels \textbf{y}, the \ac{nDCG} is defined as: 

\begin{equation}
    \text{nDCG}(\pi, \textbf{y}) = 
    \frac{\text{DCG}(\pi, \textbf{y})}{\text{DCG}(\pi^*,\textbf{y})},
    \label{eq:ndcg_general}
\end{equation}
where $\pi^*$ is the ideal ranking of the items according to the relevance labels \textbf{y}. 
The \ac{DCG} is computed as~\cite{chen2009ranking}:
\begin{equation}
\text{DCG}(\pi,\textbf{y}) = \sum_{i=1}^n \frac{2^{y_i} - 1}{\log_2(1+\text{r}(x_i, \pi))},
\end{equation}
where $\text{r}(x_i, \pi)$ is the function returning the rank of item $x_i$ in $\pi$ and $y_i$ is its label. This measure, however, is not differentiable because the function $\text{r}(\cdot, \cdot)$ is not.
To tackle this problem, Qin et al.~\cite{qin2010general} propose a differentiable approximation of \ac{nDCG} which can be used as a loss function to train \ac{ML} models~\cite{bruch2019revisiting}.
In the ApproxNDCG loss, the non-differentiable $\text{r}(\cdot, \cdot)$ function is replaced by its approximation:
\begin{equation}
    \hat{r}(x_i, \pi)=1 + \sum_{j \neq i} \mathbf{I}_{s(x_i) < s(x_j)}, \forall x_j \in \pi, 
\end{equation}
where $s(\cdot)$ indicates the scoring function of the model we are training and $\mathbf{I}_{u < v}$ is the indicator function which is 1 if $u < v$ and 0 otherwise. In turn, the indicator function is approximated with: $\sigma(v - u) = (1 + e^{-\alpha(v-u)})^{-1},$
where $\alpha$ is a parameter to control the steepness of the sigmoid function $\sigma(\cdot)$.
{
More recently, a stochastically treated version of the ApproxNDCG function was proposed by Bruch et al.~\cite{bruch2020stochastic}. In the paper, zero-mean logistic noise is added to the sigmoid function presented earlier, which becomes $\sigma((v - u) + Z_{vu})$, where $Z_{vu} \sim \text{Logistic}(\text{mean}=0, \text{scale}=\beta)$. Intuitively, adding random noise to the model outputs induces it to increase the difference between the relevance scores of relevant and not-relevant items to maintain their relative ordering in the final ranked list.
ApproxNDCG and its variant with \ac{ST} are the most popular and best performing listwise loss functions among the recently published neural \ac{LETOR} works 
such as \cite{bruch2019revisiting, bruch2020stochastic, zhuang2020interpretable, zhuang2020feature}.} 

Regarding the collection of relevance judgments, there is still an open debate in the \ac{IR} community on how to reliably collect them and on the best strategies for their aggregation. Crowdsourcing is a valuable option in this context~\cite{alonso2019practice}. The TREC Crowdsourcing track~\cite{lease2011overview, smucker2012overview, smucker2014overview} for example, explored the challenges related to the collection and management of relevance judgments for Web pages and search topics.
Numerous aggregation options are also described in the \ac{IR} literature on crowdsourcing~\cite{ravana2009score, hosseini2012aggregating, ferrante2020s} involving strategies to weigh the annotations of each judge depending on the topic difficulty and/or his/her level of confidence on it. Here, we take a new approach to this problem, eliminating the relevance judgments aggregation step and training a model on the distribution of relevance judgments associated to each document-topic pair. 

We propose five different loss functions that allow ranking models to take advantage of relevance judgments distributions prior to their aggregation. Since, to the best of our knowledge, no similar method was previously presented in the literature, we compare the newly proposed training strategies to five high-performing loss functions representative for each class of functions described before, i.e. MSE, Hinge, ApproxNDCG, ApproxNDCG with \ac{ST} and ListMLE. 

\section{Proposed Approach} \label{sec:proposed}
In this section we introduce the proposed pointwise, pairwise and listwise loss functions 
and the neural \ac{LETOR} model that we use as reference architecture.

\textit{\textbf{Pointwise Loss Functions}.}
These formulations rely on \ac{KL} divergence to compare two probability distributions, i.e. the relevance score computed by our \ac{LETOR} model for a document-topic pair and its corresponding true relevance label. 
We interpret the relevance label assigned to a document as if it was generated by a Binomial random variable modeling the judges' annotation process. For example, we assume that $n$ assessors provided one binary relevance label for each document-topic pair, i.e., to state whether the pair was a relevant or a not-relevant one. This process can be modeled as a Binomial random variable $P\sim \text{Bin}(n, p)$ where the success probability $p$ for each sample is the average of the binary responses submitted in $n$ trials. Since in most \ac{LETOR} datasets, relevance judgments are not real values in the $[0, 1]$ range, we normalize them to fit this interval. 

We then apply the same reasoning for the interpretation of our model output probability score as another Binomial distribution $\hat{P}\sim \text{Bin}(n, \hat{p})$ with the same parameter $n$ -- empirically tuned for the numerical stability of the gradients during training -- and probability $\hat{p}$ equal to the output of the model which is kept in the $[0, 1]\in \mathbb{R}$ range employing a sigmoid activation function at the output layer. 

At this point, the model output and relevance labels distributions are expressed as Binomial distributions with the same parameter $n$ and different success probabilities; hence, we can compare them using the \ac{KL} divergence:
\begin{equation}
\begin{aligned}
D_{KL_{Bin}}(P||\hat{P})= \log \left(\frac{p}{\hat{p}}\right) n p + \log \left( \frac{1-p}{1-\hat{p}}\right) n (1-p),
\end{aligned}
\label{eq:bin_kl}
\end{equation}
that is differentiable because it is the sum of two differentiable and continuous functions for $p$ and $\hat{p}$ in the open $(0, 1)$ interval. Since \ac{KL} divergence is not symmetric -- which would lead to issues in the comparison of different items during model optimization -- we employ the following symmetric and non-negative formulation as the loss function for each data point in a training batch: 
\begin{equation}
    \text{Pointwise}_{KL(Bin)} = \left(D_{KL}(P_i||\hat{P}_i) + D_{KL}(\hat{P}_i||P_i) \right) * w_i.
    \label{eq:bin_loss}
\end{equation}
Since we train our model feeding it all the items corresponding to one or more ranked lists provided in the training data, we need to balance the contributions of relevant (the minority) and not-relevant (the majority) items to the final loss function value. For this reason, before computing the total loss value in a batch by summing the contribution of each data point, we rescale each term by a factor $w_i$, inversely proportional to the number of times an item of the same class appeared in the batch. 


Since our relevance labels are graded and normalized between $[0, 1]\in \mathbb{R}$, we consider as not-relevant the data points associated to a relevance label lower than an empirically set threshold, and the remaining ones as relevant. In our experiments, we set this threshold to $0.1$ -- instead of for example $0.0$ -- for it to be used also when simulating the relevance labels distributions as described in Section \ref{sec:eval}, sampling them from continuous probability distributions.

Whenever an actual relevance judgments distribution is available, we propose to use another pointwise loss function which takes into account the distribution of values over a number of relevance grades, interpreting them as outcomes from a Multinomial distribution~\cite{bishop2006pattern, agrawal2020finite} $P\sim \text{Mul}(n, (p_1, ..., p_k))$ -- which is a generalization of the aforementioned Binomial distribution with the same parameter $n$. The outcomes of the modeled random process take a finite number of values, where $\sum_{i=1} ^k p_i = 1$. Each $p_i$ indicates the probability of one of the $k$ possible relevance grades to be selected by the pool of judges that were employed to asses a certain document-topic pair.
To obtain a comparable distribution to the output of our model we adjust the output layer size to $k$ and employ a softmax activation function over each output sequence obtaining the probabilities $(\hat{p}_1, ..., \hat{p}_k)$, which allow us to define the random variable $\hat{P}\sim \text{Mul}(n, (\hat{p}_1, ..., \hat{p}_k))$.
We then compare the two distributions with the same strategy used before, changing the formulation of the \ac{KL} divergence to Multinomial distributions:
\begin{equation}
D_{KL(Mul)}(P||\hat{P}) = \sum_{i=j}^k p_j \log\frac{p_j}{\hat{p}_j}.
\end{equation}
This function is continuous and differentiable for non-zero probability values but, again, not symmetric. Therefore, we train our model using the following symmetric and non-negative formulation to evaluate the quality of approximation of the relevance of each item:
\begin{equation}
    \text{Pointwise}_{KL(Mul)} = \left(D_{KL(Mul)}(P_i||\hat{P}_i) + D_{KL(Mul)}(\hat{P}_i||P_i) \right)*w_i.    
\label{eq:ml_loss}
\end{equation}
As in the previous case, before computing the total loss in a batch, we rescale the contribution of each data point in it by a factor $w_i$. 

{
In this case, we recommend the collection of a number of relevance labels for each query-document pair sufficient to estimate the distribution of opinions on its relevance for the user base of the search system. For example, if we are evaluating the relevance of a recently published academic study with respect to a certain topic, if our audience is made of experts from the same academic field, then a wider range of relevance labels should be collected to capture all of the nuances of the opinions field professionals could have on the topic. On the other hand, if our audience is the general public, the number of relevance judgements required can be smaller and proportional to the public agreement on the specific topic.\\
Note also that, despite the interpretation we provide for the parameter $n$ in the Binomial and Multinomial distributions as the number of judges providing relevance labels for the same query-document pair, in the remainder of this paper we consider $n$ as a hyperparameter of the model. Indeed, even if there could be a relation between the number of judges providing relevance labels and the corresponding random variable which could model this process, this intuition would be hard to verify empirically. Indeed, the number of judges available for the creation of a new \ac{IR} collection is often limited by real-world constraints -- such as labor cost or the availability of an appropriate number of trained professionals -- therefore, our hypothesis becomes hard to verify for a large number of topics where potentially dozens or even hundreds of judges could be required for the annotation of each query-document pair in an exhaustive study. 
}
\textit{\textbf{Pairwise Loss Functions}.}
We also propose two pairwise loss functions. 
The intuition here is to increase the model robustness and generalization power 
through a comparison between distributions instead of real-valued relevance scores. To achieve this goal, given a topic, we compare all pairs of relevant and not-relevant -- or less relevant if relevance labels are not binary -- documents in a batch by considering the output relevance scores produced by our model and the respective relevance labels.
{
In other words, for each topic in an experimental collection we consider every pair of documents available with a different relevance label -- obtained by either aggregating relevance scores if relevance judgments distributions are available, or taking their exact value -- and compare them to train a \ac{LETOR} model with one of the loss functions described below. }
We evaluate two possible interpretations for these relevance scores. 

The first option, similarly to what we did for the $\text{Pointwise}_{KL(Bin)}$ loss function, is to consider the relevance scores as the success probabilities of Binomial random variables and then to compute their \ac{KL} divergence.  

The second option is to consider the relevance scores of each pair as samples from two different Gaussian random variables $P^+\sim \mathcal{N}(\mu_{p^{+}}, \sigma)$ and $P^-\sim \mathcal{N}(\mu_{p^{-}}, \sigma)$ with the same standard deviation $\sigma$ but centered on the relevance labels -- or model output scores -- $\mu_{p^{+}}$ and $\mu_{p^{-}}$. 
{
Our hypothesis is that relevance judgments provided by different judges for a certain document-topic pair (relevant or not-relevant) will have a certain standard deviation ($\sigma$) and will be all centered around a certain value $\mu$, following a Gaussian random process. 
}
Therefore, if we assume the standard deviation -- i.e. the disagreement of different judges over each annotated sample -- to be constant over time, we can model the process with a Gaussian random variable with $\mu$ equal to the output relevance score of our model -- interpreted as a sample from the true distribution -- and standard deviation $\sigma$ to be adjusted according to the level of agreement/disagreement that we hypothesize in the annotation process.

{
Depending on the modeling strategy, the proposed loss functions take the following formulations typical of pairwise hinge losses~\cite{chen2009ranking}, where we replace the term dedicated to compare a pair of item with their sign-corrected KL divergence value:}
\begin{equation}
    \text{Pairwise}_{KL(Bin)} = \text{max}(0, m-\text{sign}(p^+ - p^-) D_{KL(Bin)}(P_{Bin}^+, P_{Bin}^-),
\label{eq:pw_bin_loss}
\end{equation}
{
where $m$ is a slack parameter to adjust the distance between the two distributions,  $p^+$ and $p^-$ are the outputs of the \ac{LETOR} model associated to two documents -- the former with a higher relevance label than the latter -- $P_{Bin}^+\sim\text{Bin}(n, p^+)$ and $P_{Bin}^-\sim\text{Bin}(n, p^-)$ are two Binomial} distributions corresponding to a relevant and to a not-relevant document-topic pair, respectively. The respective success probabilities are equal to the sigmoid-bounded relevance probability scores returned by the \ac{LETOR} model to train, as done for the corresponding pointwise loss function, see eq. (\ref{eq:bin_loss}). In this case, if the relevant data point has a relevance probability $p^+ > p^- $, the loss is equal to the difference between $m$ and the \ac{KL} divergence between the distributions. This difference is lower bounded by zero, thanks to the $\text{max}$ operation implemented with the \ac{ReLU} function. In this situation, the slack variable $m$ can be tuned to increase or decrease the distance between the two distributions. Conversely, if $p^- > p^+$, then the loss is positive and equal to the sum of $m$ and the value of the \ac{KL} divergence between the two distributions.

The formulation of the pairwise loss function relying on Gaussian distributions has a similar form, but it uses the formulation of the \ac{KL} divergence between two Gaussian random variables $P^+\sim\mathcal{N}(\mu_{p^{+}}, \sigma_{p^{+}})$ and $P^-\sim\mathcal{N}(\mu_{p^{-}}, \sigma_{p^{-}})$:
\begin{equation}
D_{KL(\mathcal{N})}(P||P^-)=\frac{1}{2}\left[2\log \frac{\sigma_{p^{-}}}{\sigma_{p^{+}}} - 1 + \frac{\sigma_{p^{+}}^2+(\mu_{p^{+}} - \mu_{p^{-}})^2}{\sigma_{p^{-}}^2}\right].
\label{eq:kl_gauss}
\end{equation}
This function is differentiable and, if the two random variables have the same variance, also symmetric and not-negative -- i.e. $D_{KL(\mathcal{N})}(P||P^-) = \frac{\sigma^2 + (\mu_{p^{+}} - \mu_{p^{-}})^2}{2\sigma^2} - \frac{1}{2}$. Therefore, we can employ it to train the model as: 
\begin{equation}
 \text{Pairwise}_{KL(\mathcal{N})} = \text{max}(0, m-\text{sign}(p^+ - p^-) D_{KL(\mathcal{N})}(P_{\mathcal{N}}^+, P_{\mathcal{N}}^-).
 \label{eq:pairwise2}
\end{equation}

\textit{\textbf{Listwise Loss Function}.}
{
Finally, we propose a listwise loss function also based on the \ac{KL} divergence between distributions. In this case, we consider the whole set of relevance probabilities associated to $k$ documents in a ranked lists of a batch and their respective relevance labels as two multivariate Gaussian distributions. In a similar way as in eq. (\ref{eq:kl_gauss}), we compute the \ac{KL} divergence between two multivariate Gaussian distributions $\hat{P} \sim \mathcal{N}(\pmb{\mu_{\hat{p}}}, \Sigma_{\hat{p}})$ and $P\sim \mathcal{N}(\pmb{\mu_{p}}, \Sigma_{p})$, obtaining:
\begin{equation}
\small
    D_{KL(\mathcal{N}_{mult})} = \frac{1}{2}\left[ \vphantom{\int} \text{tr}\left({\Sigma_{p}}^{-1} \Sigma_{\hat{p}}\right) + (\pmb{\mu_{p}} - \pmb{\mu_{\hat{p}}})^{T}\Sigma_{p}^{-1}(\pmb{\mu_{p}} - \pmb{\mu_{\hat{p}}})  \right. \left. - k  +  \log \left(\frac{\det\Sigma_{p}}{\det\Sigma_{\hat{p}}}\right)  \vphantom{\int} \right].
\label{eq:kl_gauss_mult}
\end{equation}}
If the two distributions have the same diagonal covariance matrix, eq. (\ref{eq:kl_gauss_mult}) reduces to: 
\begin{equation}
    D_{KL(\mathcal{N}_{mult})} = \frac{1}{2} (\pmb{\mu_{p}} - \pmb{\mu_{\hat{p}}})^T \Sigma_{p}^{-1} \left[(\pmb{\mu_{p}} - \pmb{\mu_{\hat{p}}})\right],
\end{equation}
which is also symmetric and not negative. As for the $\text{Pointwise}_{KL(Bin)}$ and $\text{Pointwise}_{KL(Mul)}$, we rescale the components of the distributions associated to relevant and not-relevant items by a factor $\pmb{w}$ inversely proportional to their respective class frequency in the current ranked list.
Therefore, the loss function value associated to one ranked list in a batch is computed as 
\begin{equation}
\text{Listwise}_{KL(\mathcal{N}_{mult})} =  D_{KL(\mathcal{N}_{mult})}(\hat{P}||P) * \pmb{w}
\label{eq:gauss_listwise_loss}
\end{equation}
where $\hat{P}\sim \mathcal{N}(\mu_{\hat{p}}, \Sigma)$ represents the output distribution of relevance probabilities returned by our model and $P\sim \mathcal{N}(\mu_{p}, \Sigma)$ indicates the distribution of true relevance judgments associated to each item in a ranked list. The loss function values associated to each ranked list are then averaged over a batch.

\begin{figure}[h!]
     \centering
     \includegraphics[scale=0.9]{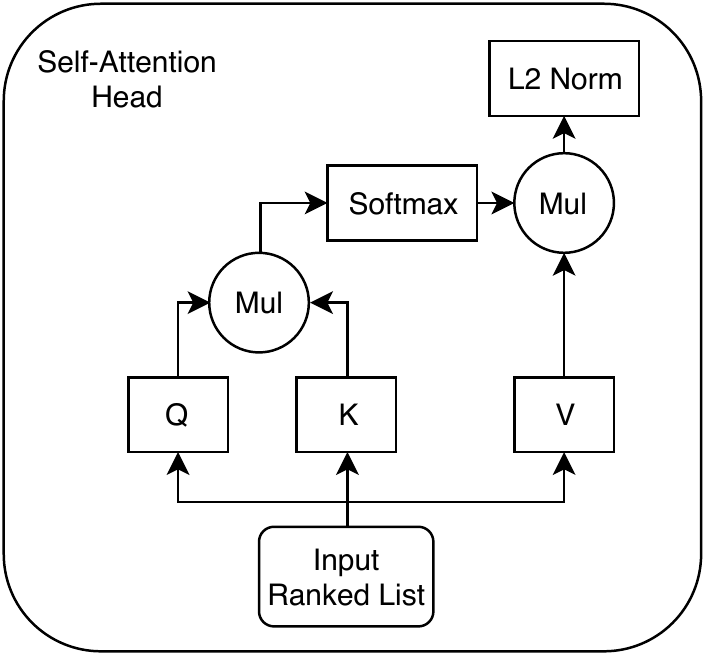}
     \caption{{Schema of the self-attention layer we employ in the proposed transformer-based \ac{LETOR} model.}}
     \label{fig:sa_l}
 \end{figure}

\textit{\textbf{Neural \ac{LETOR} Model}.}
{
The architecture of the transformer-based neural model that we employ in our experiments is depicted in Figure~\ref{fig:n_model} and is inspired to other popular ones such as~\cite{pobrotyn2020context, sun2020modeling, qin2021neural}. 
\begin{figure}[h!]
     \centering
     \includegraphics[scale=0.9]{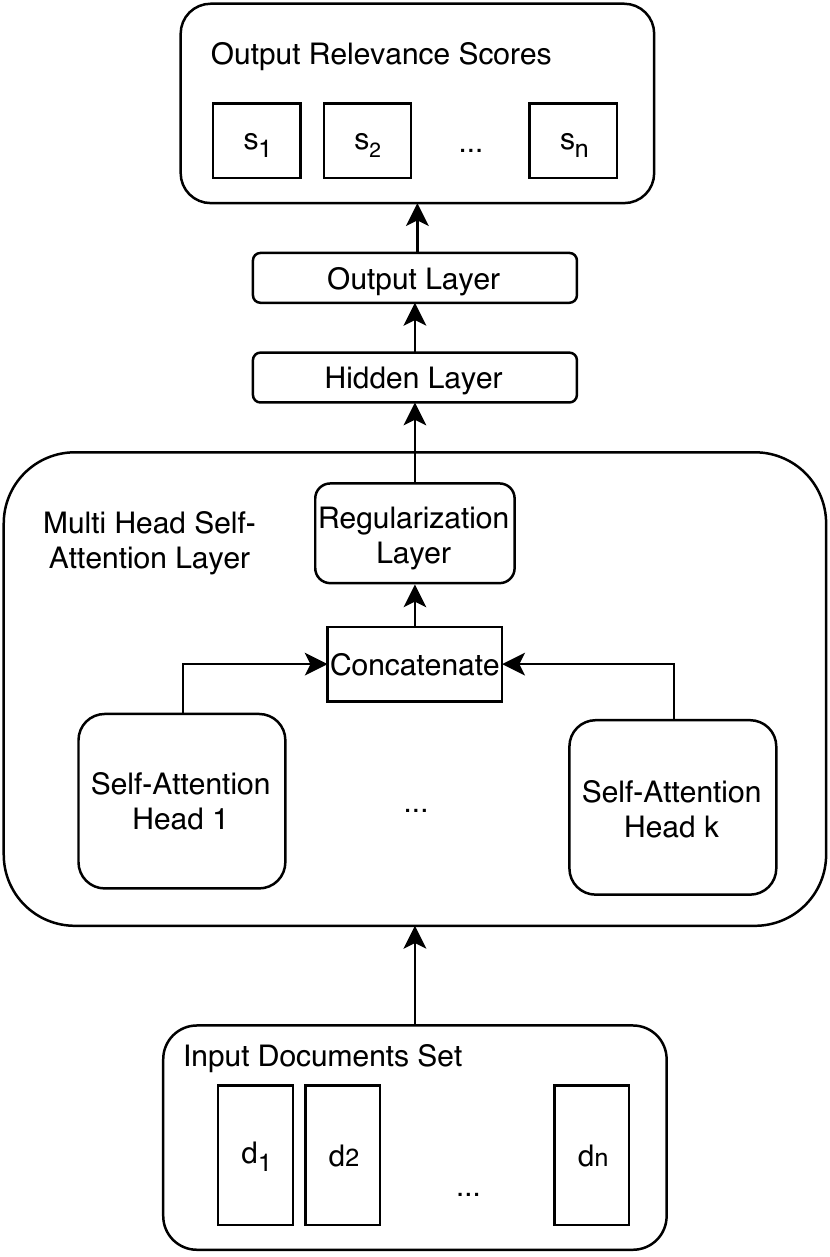}
     \caption{{Schema of the neural model employed in our experiments.}}
     \label{fig:n_model}
 \end{figure}
The first layer of the proposed architecture -- depicted in Figure~\ref{fig:sa_l} -- is a standard Multi-Head Self-Attention (SA) Layer as the one introduced in~\cite{vaswani2017attention} to compare the different document representations in input.
}
The outputs of each SA head -- of size $m=f/k$, i.e. equal to the original document representation size $f$ divided by the number $k$ of attention heads used -- are then concatenated and fed to a Regularization Layer with the goal of eliminating the redundancy in the representations of different attention heads and extracting only the features and their combinations which are useful for our goal. 
The Regularization Layer that we employ takes as input the concatenation of the outputs of different attention heads associated to the documents and begins by normalizing their components to have mean 0 and standard deviation equal to 1. 
\begin{figure}[h!]
    \centering
    \includegraphics[scale=0.7]{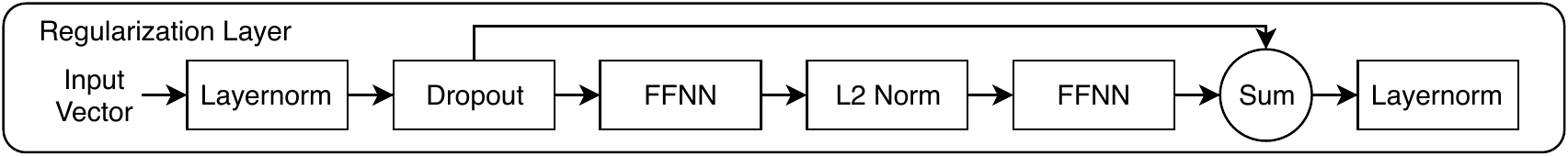}
    \caption{{Schema of the proposed Regularization Layer that we use in our \ac{LETOR} model.}}
    \label{fig:r_layer}
\end{figure}
Then, we feed these normalized representations into a \ac{FFNN} of size $t$ times the input vector size $f$, using the \ac{ReLU} activation function. Next, we normalize its output -- with the goal of forcing to zero the components of the output which are redundant -- and feed them to another \ac{FFNN} of size $f$. The output of the latter layer is then summed to the normalized input to the layer as described in Fig. \ref{fig:r_layer}. Finally, we normalize the components of the output vector of this layer to have 0 mean and unit standard deviation.
Each normalization layer estimates the components mean and standard deviations considering each feature separately. This layer is used to improve the numerical stability of the operations in the model by maintaining the values within a constant range.

The representation of each document returned by our Regularization Layer is then fed to a hidden layer of size $h$ with a \ac{ReLU} activation function and to an output layer. The output layer size is equal to the number of relevance levels of the current collection if we are using the $\text{Pointwise}_{KL(Mul)}$ loss function, one otherwise. The activation function we use depends on the output size and is the $softmax$ function if we are considering a multi-class output and the $sigmoid$ in the other cases.
\section{Experimental Setup} \label{sec:setup}

In this section, we describe the main technical details of our experimental setup, additional details are available in our code repository, along with the source code of the proposed approaches and the newly created test collection: \texttt{\url{https://github.com/albpurpura/PLTR}}.

\textit{\textbf{Experimental Collections}.}
The experimental collections that we consider in our experiments are: MQ2007, MQ2008, MSLR-WEB30K~\cite{qin2013introducing} and OHSUMED~\cite{qin2010letor}. 
All collections are already organized in five different folds with the respective training, test and validation subsets. We report the performance of our model averaged over these folds with the exception of the MSLR-WEB30K collection where we only consider Fold 1 as in other popular research works~\cite{zhuang2020feature, zhuang2020interpretable, ibrahim2016comparing, qin2021neural}. 

The MQ2007 and MQ2008 collections 
contain $1,700$ and $800$ queries, with a respective average of $40$ and $18$ assessed documents per query. 
Their relevance scores are integer values ranging from 0 to 2, indicating an increasing degree of relevance.
The MSLR-WEB30K collection is a subset of $30,000$ queries from the retired training set of the commercial search engine Microsoft Bing with an average number of $125$ judged documents per query. 
Since these features $x_i$ are not normalized, we normalize them applying the following transformation proposed in~\cite{zhuang2020feature} and also used in \cite{qin2021neural}: $\hat{x_i} = \log(|1 + x_i|) * \text{sign}(x_i)$. Relevance labels here are integers from $0$ to $4$. 
The OHSUMED~\cite{qin2010letor} collection contains about 16K documents from MEDLINE and $106$ queries with an average of $125$ assessed documents. 
Relevance labels here are in the $\{0, 1, 2\}$ set.

\textit{\textbf{Crowdsourcing Relevance judgments}.}
We also conducted a crowdsourcing experiment to obtain relevance judgments distributions for a subset of document-topic pairs from the COVID-19 MLIA collection. We consider the English MLIA subset, which contains 30 topics related to the COVID-19 pandemic and documents scraped from different online sources.~\footnote{\url{http://eval.covid19-mlia.eu}} We collected relevance judgments on an average of $64$ documents per topic (standard deviation $=4.37$) with an average of $4$ different judges (standard deviation $=7.12$) per document and with an average pairwise inter-annotator agreement of 70\%. Judges are voluntary master or Ph.D. students in computer engineering or foreign languages. The relevance labels that we consider are in the $\{0,1,2\}$ set, indicating increasing relevance. We built a \ac{LETOR} collection computing the documents features by employing different retrieval models and configurations available in the Apache Lucene framework.~\footnote{\url{https://lucene.apache.org}} We considered $24$ different retrieval pipelines using a combination of one component from each the following sets: \{BM25, Language Model with Dirichlet Smoothing (LMD), Divergence From Randomness
\} retrieval model, \{Lucene, Indri, Atire, Okapi\} stoplist -- available online~\footnote{\url{https://bitbucket.org/frrncl/gopal/src/master/src/main/resources/}} -- and \{Porter, Lovins\} stemmer.

When required by the loss functions, we aggregated the relevance judgments distributions to obtain a real-valued relevance score. This was achieved by computing a weighted average with the following weights $[-1.0, 0.5, 1.0]$, obtaining a score within $-1$ and $1$ which was then rescaled and shifted between $0$ and $1$ to be used for training and evaluation purposes. We chose this aggregation strategy as an alternative to the Majority Vote method, to better preserve all the possible degrees of relevance for a document-topic pair. 

\textit{\textbf{Model Hyperparameters and Training.}}
The training parameters of the aforementioned neural model are the number of attention heads $k$, the factor $t$ which is used to determine the size $f \times t$ of the first \ac{FFNN} in the proposed Regularization Layer, and the hidden layer size $h$. These parameters were adjusted according to the experimental collection size and number of features of the input data points and tuned on the validation sets of the first folds of each collection. The model was trained for a maximum of $100$ epochs with early stopping with patience of $20$ epochs on the MQ2007, MQ2008 and MSLR-WEB30K collections, while we reduced the maximum number of training epochs to $50$ on the OHSUMED and MLIA collection. The best model was then selected for each fold according to the nDCG@1 performance on the validation set. 
The number of attention heads $k$ was dependent both on the collection size and the number of features available to represent each document. This parameter was set to $2$ for the experiments on the MQ2007 and MQ2008 collections, $4$ on the MSLR-WEB30K, $3$ on the OHSUMED and $1$ on the MLIA collection. The factor $t$ was empirically set to $3$ for the experiments on all collections, while the hidden layer size $h$ was set to $32$ for the experiments on the MQ2007, MQ2008 and OHSUMED collections, to $128$ for the ones on the MSLR-WEB30K and to $8$ for the experiments on MLIA.

Since in the MSLR-WEB30K and OHSUMED collections the maximum number of documents per query is much larger than the average on the other datasets, we reranked in these cases only the top 150 documents ranked by a LightGBM LambdaMART model~\cite{ke2017lightgbm} tuned considering the nDCG@1 measure on the validation set of each collection fold. We cut the ranked lists at $150$ documents since this was also the maximum value of documents available to rerank per-query on the MQ2007, MQ2008 and MLIA collections.


{
In our experimental section, we also provide a performance comparison of our probabilistic training approach simulating a noisy annotation process.
For each annotated query-document pair, we sample a new set of relevance labels from a Binomial distribution with parameters $n=32$ -- we determined this value empirically according of the performance of the model on the validation set of the first fold of the experimental collections in the previous experiments with pointwise loss functions --
and $p$ equal to the normalized relevance label found in the dataset.
In other words, if a dataset employs relevance labels in the set \{0, 1, 2\}, then the $p$ values we employ to sample new relevance labels will take values in the set \{0, 0.5, 1\}, in the same order. 
After the sampling step, we average the sampled values to generate a new ``sampled'' relevance label. This label is used as the parameter $p$ in equations ~\ref{eq:bin_loss}, \ref{eq:ml_loss}, \ref{eq:gauss_listwise_loss} and compared to the real-valued output of the model we train indicated as $\hat{p}$. The sampled relevance label is also used to compute which documents to consider in each training pair provided to the pairwise loss functions described in equations \ref{eq:pw_bin_loss} and \ref{eq:pairwise2}.
This process does not change the relevance label of highly relevant or not-relevant documents -- i.e. the documents with the highest or lowest relevance grade that have a success probability of $1.0$ or $0.0$, respectively -- but provides some range of variability for the documents judged as \textit{partially-relevant}, with a variance proportional to the uncertainty on the relevance of the document. Indeed the variance of each Binomial variable $P\sim \text{Bin}(n, p)$ is defined as $np(1-p)$, it is therefore higher for values closer to $0.5$ -- i.e. values of $p$ associated to relevance labels in the middle of the grading scale. 
Intuitively, the parameter $n$ increases or decreases the width of the \ac{PDF} of the distributions we compare, i.e. a larger $n$ can be used in the cases where there is a large variability in the relevance judgements distribution for the same query-document pair and vice versa. Therefore, we recommend treating $n$ as a model parameter and tuning it on a separate validation set. 
}

\textit{\textbf{Evaluation Measures and Baselines}.}
We evaluate the performance of the proposed loss functions relying on top-heavy, widely-used measures as ERR~\cite{chapelle2009expected}, nDCG@\{1, 3, 5\} \footnote{The nDCG formulation we employ is the one used by the \texttt{TREC\_eval} tool.~\url{https://github.com/usnistgov/trec_eval}} and P@\{1, 3, 5\}; these are also amongst the most used measures in the LETOR literature. Moreover, despite recent critics~\cite{FerranteFL20, Fuhr2018} and the open debate in the community, to ease the comparison with other \ac{LETOR} approaches in the literature~\cite{ tax2015cross,ibrahim2016comparing, koppel2019pairwise}, we also report the Average Precision (AP) of each of our runs averaged over all topics (MAP).
We also compute a paired Student's t-test for each measure and report the performance difference between the baseline of choice (i.e., the proposed neural model trained with the ApproxNDCG loss function or the LambdaMART model, depending on the experiment at hand) and the same model trained with the other loss functions; we indicate with 
$\uparrow$ or $\downarrow$ a statistically significant difference ($\alpha<0.05$), accordingly to the sign of the difference. 
We selected the ApproxNDCG loss as a reference loss function for the proposed neural model since it was the one with the best performance across all the considered collections. 
To simplify our experimental analysis, we report the performance of our model when trained using the following loss functions as representatives from the pointwise, pairwise and listwise loss categories: MSE, Hinge, ApproxNDCG with and without \ac{ST} and ListMLE.

Given the success of the LambdaMART model in the \ac{LETOR} domain and its widespread usage as a strong reference baseline for the evaluation of new systems~\cite{qin2021neural}, we also evaluate the impact of the proposed loss functions to train a decision tree-based \ac{GBM} model, that is the model at the base of LambdaMART~\cite{burges2010ranknet}.
We also evaluate the impact of the proposed training strategies on two popular neural \ac{LETOR} models, i.e. \ac{DASALC}~\cite{qin2021neural} and a simple transformer model with one self attention layer similar to the one used in ~\cite{pobrotyn2020context}.

To perform the experiments with the \ac{GBM} model, we rely on the LightGBM library. On each collection, we tune the hyperparameters of the \ac{GBM} to obtain the highest nDCG@1 on the validation set of Fold 1 of each experimental collection. The hyperparameter optimization process is performed -- similarly to ~\cite{qin2013introducing} -- through a grid search over the following parameters: learning rate $\{0.001, 0.05, 0.1, 0.5\}$, number of trees $\{300, 500, 1000\}$, number of leaves $\{200, 500, 1000\}$ and followed by an additional manual tuning around the best hyperparameters combination found in the previous step. The best model hyperparameters for each collection are reported in our online repository.
{
We also compare the performance of the neural \ac{LETOR} and \ac{GBM} model to the LightGBM LambdaMART implementation and to other neural models. We tuned the hyperparameters of each neural model we consider following the same criteria described in~\cite{qin2021neural}. 
}

\section{Evaluation} \label{sec:eval}
\begin{table*}[t!] \centering
\resizebox{\textwidth}{!}{\begin{tabular}{@{\hskip3pt}l@{\hskip3pt}@{\hskip3pt}l@{\hskip3pt}llllllll} & Loss Function & ERR & P@1 & P@3 & P@5 & nDCG@1 & nDCG@3 & nDCG@5 & AP \\ \hline
\multirow{8}{*}{MQ2007}
& \underline{ApproxNDCG} & 0.3169 & 0.4639 & 0.4291 & 0.4105 & 0.4152 & 0.4150 & 0.4219 & 0.4603 \\
& ListMLE & 0.3178  & 0.4681  & 0.4336 & 0.4116 & 0.4178 & 0.4175 & 0.4215 & 0.4596 \\
& ApproxNDCG (ST) & 0.2704$\downarrow$  & 0.3842$\downarrow$ & 0.3621$\downarrow$  & 0.3494$\downarrow$ & 0.3363$\downarrow$  & 0.3391$\downarrow$ & 0.3459$\downarrow$ & 0.4021$\downarrow$ \\
& Hinge & 0.2674$\downarrow$ & 0.3723$\downarrow$ & 0.3633$\downarrow$  & 0.3511$\downarrow$ & 0.3239$\downarrow$ & 0.3334$\downarrow$& 0.3420$\downarrow$ & 0.4004$\downarrow$ \\
& MSE & 0.3154 & 0.4574 & 0.4291  & 0.4115  & 0.4113  & 0.4136  & 0.4205  & 0.4550$\downarrow$ \\ \cdashline{2-10}
& Pointwise KL (Binomial) & 0.3168 & 0.4551 & 0.4334 & 0.4132  & 0.4087 & 0.4177 & 0.4230 & 0.4601 \\ 
& Pairwise KL (Binomial) & 0.3196 & 0.4728 & 0.4377 & 0.4186$\uparrow$ & 0.4249 & 0.4226 & 0.4297$\uparrow$ & 0.4647 \\
& \textbf{Pairwise KL (Gaussian)} & \textbf{0.3218} & \textbf{0.4817}$\uparrow$ & \textbf{0.4381} & \textbf{0.4201}$\uparrow$ & \textbf{0.4350}$\uparrow$ & \textbf{0.4249}$\uparrow$ & \textbf{0.4318}$\uparrow$ & \textbf{0.4665}$\uparrow$ \\
& Listwise KL (Gaussian) & 0.3177 & 0.4657 & 0.4332 & 0.4145 & 0.4173 & 0.4192 & 0.4255 & 0.4634 \\\hline\hline
\multirow{8}{*}{MQ2008}
& \underline{ApproxNDCG} & 0.2972 & 0.4222 & 0.3639 & 0.3337 & 0.3750 & 0.4039 & 0.4484 & 0.4585 \\
& ListMLE & 0.2801$\downarrow$ & 0.3954$\downarrow$ & 0.3482$\downarrow$ & 0.3153$\downarrow$ & 0.3508$\downarrow$ & 0.3795$\downarrow$ & 0.4236$\downarrow$ & 0.4399$\downarrow$ \\
& ApproxNDCG (ST) & 0.2783$\downarrow$ & 0.4005 & 0.3571 & 0.3171$\downarrow$ & 0.3457$\downarrow$ & 0.3856 & 0.4234$\downarrow$ & 0.4373$\downarrow$ \\
& Hinge & 0.2642$\downarrow$ & 0.3533$\downarrow$ & 0.3350$\downarrow$ & 0.3099$\downarrow$ & 0.3087$\downarrow$ & 0.3548$\downarrow$ & 0.4035$\downarrow$ & 0.4228$\downarrow$ \\
& MSE & 0.2997 & \textbf{0.4401} & 0.3844$\uparrow$ & 0.3431$\uparrow$ & 0.3795 & 0.4177$\uparrow$ & 0.4596$\uparrow$ & 0.4709$\uparrow$ \\\cdashline{2-10}
& Pointwise KL (Binomial) & 0.2968 & 0.4222 & 0.3797$\uparrow$ & 0.3375 & 0.3699 & 0.4126 & 0.4520 & 0.4626 \\
& Pairwise KL (Binomial) & 0.2995 & 0.4388 & 0.3797$\uparrow$ & 0.3375 & 0.3839  & 0.4179 & 0.4567 & 0.4681$\uparrow$ \\
& \textbf{Pairwise KL (Gaussian)} & \textbf{0.3019} & 0.4375 & \textbf{0.3852}$\uparrow$ & 0.3398 & \textbf{0.3871} & \textbf{0.4222}$\uparrow$ & 0.4603$\uparrow$ & 0.4697$\uparrow$ \\
& Listwise KL (Gaussian) & 0.3008 & 0.4349 & 0.3814$\uparrow$ & \textbf{0.3457}$\uparrow$ & 0.3827 & 0.4171 & \textbf{0.4630}$\uparrow$ & \textbf{0.4729}$\uparrow$  \\\hline\hline
\multirow{8}{*}{WEB30K}
& \underline{ApproxNDCG} & 0.3523 & 0.7504 & 0.7158 & 0.6936 & 0.5263 & 0.5090 & 0.5084 & 0.5798 \\
& ListMLE & 0.3354$\downarrow$ & 0.7756$\uparrow$ & 0.7362$\uparrow$ & 0.7103$\uparrow$ & 0.5217 & 0.5066  & 0.5068 & \textbf{0.5983}$\uparrow$ \\
& ApproxNDCG (ST) & 0.3707$\uparrow$ & 0.7804$\uparrow$ & 0.6994$\downarrow$  & 0.6811$\downarrow$ & 0.5769$\uparrow$ & 0.5065 & 0.5040$\downarrow$ & 0.5818$\uparrow$ \\
& \textbf{Hinge} & \textbf{0.3748}$\uparrow$ & \textbf{0.7885}$\uparrow$ & 0.6899$\downarrow$ & 0.6745$\downarrow$ & \textbf{0.5884}$\uparrow$ & 0.5054  & 0.5046  & 0.5877$\uparrow$  \\
& MSE & 0.3623$\uparrow$  & 0.7709$\uparrow$  & 0.7334$\uparrow$  & 0.7093$\uparrow$  & 0.5461$\uparrow$  & \textbf{0.5268}$\uparrow$  & \textbf{0.5248}$\uparrow$  & 0.5921$\uparrow$  \\\cdashline{2-10}
& Pointwise KL (Binomial) & 0.3502  & 0.7596  & 0.7271$\uparrow$  & 0.6985$\uparrow$  & 0.5293  & 0.5136$\uparrow$  & 0.5093  & 0.5862$\uparrow$  \\
& Pairwise KL (Binomial) & 0.3467$\downarrow$  & 0.7812$\uparrow$  & \textbf{0.7446}$\uparrow$  & \textbf{0.7184}$\uparrow$  & 0.5357$\uparrow$  & 0.5242$\uparrow$  & 0.5241$\uparrow$  & 0.5971$\uparrow$  \\
& Pairwise KL (Gaussian) & 0.3454$\downarrow$  & 0.7753$\uparrow$  & 0.7423$\uparrow$  & 0.7166$\uparrow$  & 0.5315  & 0.5209$\uparrow$  & 0.5214$\uparrow$  & 0.5971$\uparrow$  \\
& Listwise KL (Gaussian) & 0.3523  & 0.7612$\uparrow$  & 0.7258$\uparrow$  & 0.7016$\uparrow$  & 0.5322  & 0.5152$\uparrow$  & 0.5141$\uparrow$  & 0.5871$\uparrow$  \\\hline\hline
\multirow{8}{*}{OHSUMED}
& \underline{ApproxNDCG} & 0.4981 & 0.5660 & 0.5377 & 0.5057 & 0.4764 & 0.4544 & 0.4371 & 0.3828 \\
& ListMLE & 0.4732 & 0.5377 & 0.4717$\downarrow$ & 0.4321$\downarrow$ & 0.4575 & 0.3972$\downarrow$ & 0.3737$\downarrow$ & 0.3220$\downarrow$\\
& ApproxNDCG (ST) & 0.4159$\downarrow$  & 0.5000  & 0.4119$\downarrow$  & 0.3660$\downarrow$  & 0.4009  & 0.3330$\downarrow$  & 0.3098$\downarrow$  & 0.2882$\downarrow$  \\
& Hinge & 0.4609  & 0.5660 & 0.4748$\downarrow$  & 0.4566  & 0.4434  & 0.3930$\downarrow$  & 0.3805$\downarrow$  & 0.3530$\downarrow$  \\
& MSE & 0.4376$\downarrow$  & 0.5377  & 0.4465$\downarrow$  & 0.4057$\downarrow$  & 0.4340  & 0.3663$\downarrow$  & 0.3417$\downarrow$  & 0.3049$\downarrow$  \\\cdashline{2-10}
& Pointwise KL (Binomial) & 0.4875  & 0.5189  & 0.5031  & 0.5019  & 0.4481  & 0.4269  & 0.4312  & 0.3800  \\
& Pairwise KL (Binomial) & 0.4206$\downarrow$  & 0.4811  & 0.4151$\downarrow$  & 0.3868$\downarrow$  & 0.3915$\downarrow$  & 0.3360$\downarrow$  & 0.3235$\downarrow$  & 0.2886$\downarrow$  \\
& Pairwise KL (Gaussian) & 0.4520  & 0.5377  & 0.4403$\downarrow$  & 0.4000$\downarrow$  & 0.4481  & 0.3707$\downarrow$  & 0.3481$\downarrow$  & 0.2903$\downarrow$  \\
& \textbf{Listwise KL (Gaussian)} & \textbf{0.5248}  & \textbf{0.6038}  & \textbf{0.5692}  & \textbf{0.5057}  & \textbf{0.5189}  & \textbf{0.4796}  & \textbf{0.4456}  & \textbf{0.3861}  \\\hline
\end{tabular}}

\caption{Performance of the proposed \ac{LETOR} neural model averaged over all topics. $\uparrow$ or $\downarrow$ indicate a statistically significant difference ($\alpha<0.05$) with the performance obtained using the ApproxNDCG loss function. Best performance measures per collection are in bold as the loss function with the majority of best measures per collection.}
\label{tab:sim_labels=false}
\end{table*}

In Table~\ref{tab:sim_labels=false}, we report the performance of the proposed \ac{LETOR} neural model trained using different loss functions on the MQ2007, MQ2008, MSLR-WEB30K and OHSUMED collections. 
We observe that the neural model achieves the best performance in the majority of the collections when trained with the $\text{Pairwise}_{KL(\mathcal{N})}$ loss function. On the MQ2007 and MQ2008 collections, all the proposed pairwise and listwise functions outperform the ApproxNDCG loss and most of the other losses with the exception of the MSE loss on the MQ2008 collection which is the most competitive amongst the baselines. On the MSLR-WEB30K collection the best of the proposed $\text{Pairwise}_{KL(Bin)}$ loss functions outperforms the best loss function overall on this collection, i.e. the Hinge loss, in terms of P@3, P@5, nDCG@3, nDCG@5 and AP. We also observe that the performance of the model trained using the ApproxNDCG with \ac{ST} loss is higher here compared to other collections. Indeed, on smaller collections the amount of training data is probably not sufficient for the model to benefit from this training strategy. On the OHSUMED collection, the best loss function is the $\text{Listwise}_{KL(\mathcal{N}_{mult})}$, followed by the ApproxNDCG loss. This is due to the combination of a high proportion of relevant documents in the ranked lists and little amount of training data. For all the experiments on this collection, we rerank the top 150 documents returned by a LambdaMART model and, for this reason, we expect the representations of the documents in each ranked list to be more similar to each other rather than in other collections. Hence, approaches that consider multiple documents at a time, might have an advantage in finding the differences between them and providing more insightful information through their gradients to the model during training.
{

\begin{table*}[h!] \centering
\resizebox{\textwidth}{!}{\begin{tabular}{@{\hskip3pt}p{2cm}@{\hskip3pt}@{\hskip3pt}l@{\hskip3pt}llllllll} & Loss Function & ERR & P@1 & P@3 & P@5 & nDCG@1 & nDCG@3 & nDCG@5 & AP \\ \hline
\multirow{8}{*}{MQ2007}
& \underline{ApproxNDCG} & 0.3175 & 0.4645 & 0.4342 & 0.4090 & 0.4125 & 0.4186 & 0.4217 & 0.4608 \\
& ListMLE & 0.3174 & 0.4639 & 0.4350 & 0.4110 & 0.4158 & 0.4176 & 0.4209 & 0.4591 \\
& ApproxNDCG (ST) & 0.2749$\downarrow$  & 0.3895$\downarrow$  & 0.3690$\downarrow$  & 0.3579$\downarrow$  & 0.3398$\downarrow$  & 0.3459$\downarrow$  & 0.3540$\downarrow$  & 0.4061$\downarrow$  \\
& Hinge & 0.2667$\downarrow$  & 0.3806$\downarrow$  & 0.3599$\downarrow$  & 0.3486$\downarrow$  & 0.3295$\downarrow$  & 0.3331$\downarrow$  & 0.3416$\downarrow$  & 0.4017$\downarrow$  \\
& MSE & 0.3162  & 0.4569  & 0.4334  & 0.4117  & 0.4122  & 0.4167  & 0.4208  & 0.4554$\downarrow$  \\\cdashline{2-10}
& Pointwise KL (Binomial) & 0.3185  & 0.4569  & 0.4297  & 0.4142  & 0.4119  & 0.4159  & 0.4246  & 0.4602  \\
& Pairwise KL (Binomial) & 0.3147  & 0.4592  & 0.4356  & 0.4158  & 0.4116  & 0.4163  & 0.4225  & 0.4612  \\
& \textbf{Pairwise KL (Gaussian)} & \textbf{0.3193}  & \textbf{0.4758}  & \textbf{0.4383}  & \textbf{0.4201}$\uparrow$  & \textbf{0.4288}$\uparrow$  & \textbf{0.4235}  & \textbf{0.4294}$\uparrow$  & \textbf{0.4655}  \\
& Listwise KL (Gaussian) & 0.3187  & 0.4616  & 0.4370  & 0.4178$\uparrow$  & 0.4146  & 0.4204  & 0.4272  & 0.4621  \\\hline\hline
\multirow{8}{*}{MQ2008}
& \underline{ApproxNDCG} & 0.2915 & 0.4120 & 0.3588 & 0.3286 & 0.3661 & 0.3951 & 0.4381 & 0.4498 \\
& ListMLE & 0.3008$\uparrow$ & 0.4413$\uparrow$ & 0.3797$\uparrow$ & 0.3362$\uparrow$ & 0.3897$\uparrow$ & 0.4186$\uparrow$  & 0.4582$\uparrow$ & 0.4694$\uparrow$ \\
& ApproxNDCG (ST) & 0.2659$\downarrow$  & 0.3890  & 0.3384$\downarrow$  & 0.3077$\downarrow$  & 0.3361  & 0.3676$\downarrow$  & 0.4091$\downarrow$  & 0.4297$\downarrow$  \\
& Hinge & 0.2606$\downarrow$  & 0.3457$\downarrow$  & 0.3316$\downarrow$  & 0.3092$\downarrow$  & 0.3061$\downarrow$  & 0.3521$\downarrow$  & 0.4005$\downarrow$  & 0.4194$\downarrow$  \\
& MSE & 0.2991  & 0.4260  & 0.3835$\uparrow$  & 0.3416$\uparrow$  & 0.3744  & 0.4135$\uparrow$  & 0.4568$\uparrow$  & 0.4663$\uparrow$  \\\cdashline{2-10}
& Pointwise KL (Binomial) & 0.2999  & 0.4311  & 0.3831$\uparrow$  & 0.3378$\uparrow$  & 0.3782  & 0.4193$\uparrow$  & 0.4554$\uparrow$  & 0.4671$\uparrow$  \\
& Pairwise KL (Binomial) & 0.3016$\uparrow$  & 0.4362$\uparrow$  & 0.3814$\uparrow$  & 0.3390$\uparrow$  & 0.3871  & 0.4173$\uparrow$  & 0.4584$\uparrow$  & 0.4667$\uparrow$  \\
& \textbf{Pairwise KL (Gaussian)} & \textbf{0.3081}$\uparrow$  & \textbf{0.4503}$\uparrow$  & \textbf{0.3946}$\uparrow$  & \textbf{0.3449}$\uparrow$  & \textbf{0.4005}$\uparrow$  & \textbf{0.4314}$\uparrow$  & \textbf{0.4680}$\uparrow$  & \textbf{0.4769}$\uparrow$  \\
& Listwise KL (Gaussian) & 0.2996  & 0.4311  & 0.3865$\uparrow$  & 0.3439$\uparrow$  & 0.3788  & 0.4210$\uparrow$  & 0.4611$\uparrow$  & 0.4717$\uparrow$  \\\hline\hline
\multirow{8}{*}{WEB30K}
& \underline{ApproxNDCG} & 0.3522 & 0.7523 & 0.7171 & 0.6938 & 0.5263 & 0.5091 & 0.5082 & 0.5797 \\
& ListMLE & 0.3336$\downarrow$  & 0.7732$\uparrow$ & 0.7352$\uparrow$ & 0.7102$\uparrow$ & 0.5181 & 0.5052 & 0.5061 & 0.5980$\uparrow$\\
& ApproxNDCG (ST) & 0.1762$\downarrow$  & 0.5103$\downarrow$  & 0.5086$\downarrow$  & 0.5060$\downarrow$  & 0.2614$\downarrow$  & 0.2773$\downarrow$  & 0.2913$\downarrow$  & 0.5041$\downarrow$  \\
& \textbf{Hinge} & \textbf{0.3853}$\uparrow$  & \textbf{0.7919}$\uparrow$  & 0.7126  & 0.6998$\uparrow$  & \textbf{0.5926}$\uparrow$  & \textbf{0.5310}$\uparrow$  & \textbf{0.5316}$\uparrow$  & 0.5887$\uparrow$  \\
& MSE & 0.3638$\uparrow$  & 0.7718$\uparrow$  & 0.7344$\uparrow$  & 0.7098$\uparrow$  & 0.5483$\uparrow$  & 0.5286$\uparrow$  & 0.5262$\uparrow$  & 0.5924$\uparrow$  \\ \cdashline{2-10}
& Pointwise KL (Binomial) & 0.3545  & 0.7542  & 0.7245$\uparrow$  & 0.6986$\uparrow$  & 0.5305  & 0.5137$\uparrow$  & 0.5106  & 0.5883$\uparrow$  \\
& Pairwise KL (Binomial) & 0.3475$\downarrow$  & 0.7777$\uparrow$  & 0.7406$\uparrow$  & 0.7150$\uparrow$  & 0.5352$\uparrow$  & 0.5208$\uparrow$  & 0.5212$\uparrow$  & 0.5957$\uparrow$  \\
& Pairwise KL (Gaussian) & 0.3506  & 0.7812$\uparrow$  & \textbf{0.7475}$\uparrow$  & \textbf{0.7202}$\uparrow$  & 0.5402$\uparrow$  & 0.5272$\uparrow$  & 0.5267$\uparrow$  & \textbf{0.5988}$\uparrow$  \\
& Listwise KL (Gaussian) & 0.3513  & 0.7529  & 0.7234$\uparrow$  & 0.6987$\uparrow$  & 0.5250  & 0.5129  & 0.5111  & 0.5870$\uparrow$  \\\hline\hline
\multirow{8}{*}{OHSUMED}
& \underline{ApproxNDCG} & 0.5157 & \textbf{0.6038} & \textbf{0.5377} & 0.4906 & \textbf{0.5189} & \textbf{0.4684} & 0.4395 & 0.3729 \\
& ListMLE & 0.4616$\downarrow$ & 0.5377 & 0.4748$\downarrow$ & 0.4151$\downarrow$ & 0.4387$\downarrow$ & 0.4002$\downarrow$ & 0.3667$\downarrow$ & 0.3116$\downarrow$ \\
& ApproxNDCG (ST) & 0.4584$\downarrow$  & 0.5660  & 0.4465$\downarrow$  & 0.4208$\downarrow$  & 0.4623  & 0.3765$\downarrow$  & 0.3613$\downarrow$  & 0.3257$\downarrow$  \\
& Hinge & 0.4278$\downarrow$  & 0.5094$\downarrow$  & 0.4465$\downarrow$  & 0.4358$\downarrow$  & 0.4009$\downarrow$  & 0.3604$\downarrow$  & 0.3517$\downarrow$  & 0.3475$\downarrow$  \\
& MSE & 0.4474$\downarrow$  & 0.5566  & 0.4528$\downarrow$  & 0.4075$\downarrow$  & 0.4434$\downarrow$  & 0.3776$\downarrow$  & 0.3493$\downarrow$  & 0.3160$\downarrow$  \\ \cdashline{2-10}
& Pointwise KL (Binomial) & 0.4820  & 0.5094  & 0.5000  & 0.4830  & 0.4340  & 0.4244  & 0.4177  & 0.3727  \\
& Pairwise KL (Binomial) & 0.4203$\downarrow$  & 0.4717$\downarrow$  & 0.4182$\downarrow$  & 0.3811$\downarrow$  & 0.3915$\downarrow$  & 0.3376$\downarrow$  & 0.3185$\downarrow$  & 0.2887$\downarrow$  \\
& Pairwise KL (Gaussian) & 0.4446$\downarrow$  & 0.5094$\downarrow$  & 0.4340$\downarrow$  & 0.3906$\downarrow$  & 0.4245$\downarrow$  & 0.3647$\downarrow$  & 0.3401$\downarrow$  & 0.2933$\downarrow$  \\
& \textbf{Listwise KL (Gaussian)} & \textbf{0.5165}  & 0.5755  & \textbf{0.5377} & \textbf{0.5113}  & 0.5000  & 0.4654  & \textbf{0.4473}  & \textbf{0.3873}$\uparrow$  \\\hline
\end{tabular}}\caption{Performance of the proposed \ac{LETOR} neural model averaged over all topics. The model is trained sampling the relevance labels from a Binomial distribution. $\uparrow$ or $\downarrow$ indicate a statistically significant difference ($\alpha<0.05$) with the ApproxNDCG baseline. Best performance measures per collection are in bold as the loss function with the majority of best measures per collection.}
\label{tab:sim_labels=true}
\end{table*}

In Table \ref{tab:sim_labels=true}, we report the results of the experiments training the proposed \ac{LETOR} neural model using sampled relevance judgments as described in Section \ref{sec:setup}. 
As we can see in this table, the relative performance of the neural model when relying on different loss functions remains similar and the best loss function overall is still the $\text{Pairwise}_{KL(\mathcal{N})}$ loss. However, the neural model performance is often higher in this case than in the previous experimental setup, regardless of the loss function used. This is true for at least one performance measure when using all but the $\text{Pairwise}_{KL(\text{Bin})}$ loss on the MQ2007, and for all the proposed losses on the MQ2008 collection.
We also observe a few performance improvements in the MSLR-WEB30K collection when using all loss functions with the exception of the ApproxNDCG with \ac{ST} and the ListMLE loss. Finally, on the OHSUMED collection, we observe a performance improvement in at least one measure when using the ApproxNDCG, ApproxNDCG with \ac{ST}, MSE or $\text{Pairwise}_{KL(\text{Bin})}$ loss. These results support our hypothesis that acknowledging and exploiting the possible inconsistencies in the training data can be a viable way to improve a \ac{LETOR} model's performance.

\begin{table*}[h!] \centering \resizebox{\textwidth}{!}{\begin{tabular}{@{\hskip3pt}l@{\hskip3pt}@{\hskip3pt}l@{\hskip3pt}llllllll} & Loss Function & ERR & P@1 & P@3 & P@5 & nDCG@1 & nDCG@3 & nDCG@5 & AP\\ \hline
\multirow{9}{*}{MLIA}
& \underline{ApproxNDCG} & 0.3313 & 0.4000 & 0.4667 & 0.4800 & 0.3556 & 0.3896 & 0.3973 & 0.3593 \\
& ApproxNDCG (ST) & 0.2544  & 0.3000  & 0.3333$\downarrow$  & 0.3267$\downarrow$  & 0.2708  & 0.2788  & 0.2820$\downarrow$  & 0.3144  \\
& ListMLE & 0.2463$\downarrow$ & 0.2667 & 0.3111$\downarrow$ & 0.3133$\downarrow$ & 0.2387 & 0.2640 & 0.2705$\downarrow$ & 0.3162 \\
& Hinge & 0.3455  & \textbf{0.5000}  & 0.4667  & 0.4467  & \textbf{0.4330}  & 0.3830  & 0.3736  & 0.3822  \\
& MSE & 0.2675  & 0.3333  & 0.3333  & 0.3333$\downarrow$  & 0.2917  & 0.2894  & 0.2900$\downarrow$  & 0.3101$\downarrow$  \\ \cdashline{2-10}
& \textbf{Pointwise KL (Multinomial)} & \textbf{0.3377}  & 0.4333  & \textbf{0.5000}  & \textbf{0.4933}  & 0.3628  & \textbf{0.4096}  & \textbf{0.4054}  & \textbf{0.3834}$\uparrow$  \\
& Pointwise KL (Binomial) & 0.2562  & 0.3333  & 0.3222$\downarrow$  & 0.3533$\downarrow$  & 0.2798  & 0.2684$\downarrow$  & 0.2898$\downarrow$  & 0.3122  \\
& Pairwise KL (Binomial) & 0.2639  & 0.3667  & 0.3444$\downarrow$  & 0.3200$\downarrow$  & 0.3111  & 0.2919$\downarrow$  & 0.2814$\downarrow$  & 0.3023$\downarrow$  \\
& Pairwise KL (Gaussian) & 0.2423$\downarrow$  & 0.3333  & 0.3000$\downarrow$  & 0.2867$\downarrow$  & 0.2750  & 0.2515$\downarrow$  & 0.2498$\downarrow$  & 0.3062  \\
& Listwise KL (Gaussian) & 0.2521  & 0.3333  & 0.3111$\downarrow$  & 0.3133$\downarrow$  & 0.2715  & 0.2656$\downarrow$  & 0.2653$\downarrow$  & 0.3258  \\\hline\hline
\end{tabular}}\caption{Performance of the proposed \ac{LETOR} neural model on the COVID19-MLIA collection averaged over all topics. $\uparrow$ or $\downarrow$ indicate a statistically significant difference ($\alpha<0.05$) with the ApproxNDCG baseline. Best performance measures per collection are in bold as the loss function with the majority of best measures per collection.}
\label{tab:mlia}
\end{table*}

In Table~\ref{tab:mlia}, we report the experiments of the crowdsourcing experiments on the COVID19-MLIA collection. In this case, we set the output size of the neural model to $3$ -- the same number of relevance grades that we used for our annotation -- and trained the model either (i) by aggregating -- as explained in Section \ref{sec:setup} -- the model output probability scores and the collected relevance judgments distributions in the same way, obtaining one relevance score and relevance label to train the model with the previously evaluated loss functions, or (ii) by training the model with the proposed $\text{Pointwise}_{KL(Mul)}$ loss function which can take into account the raw probability distributions over the three relevance classes. Note that, since relevance judgements distributions are only available for this collections, this is the only scenario where we can employ the proposed $\text{Pointwise}_{KL(Mul)}$ loss function.

We observe that the proposed $\text{Pointwise}_{KL(Mul)}$ loss function is the best loss function to train the neural model on this collection. This can also be partially due to the small size of the collection which favors pointwise and pairwise loss functions. However, $\text{Pointwise}_{KL(Mul)}$ still outperforms other pointwise loss functions such as the MSE and the $\text{Pointwise}_{KL(Bin)}$ losses by a sizable margin. The Hinge loss is the second best loss function on this collection, outperforming all other baselines.
The results from this experiment further confirm our initial hypothesis on the feasibility of training a \ac{LETOR} model using raw probability distributions as training data.

\begin{table*}[h!] \centering \resizebox{\textwidth}{!}{\begin{tabular}{@{\hskip3pt}l@{\hskip3pt}@{\hskip3pt}l@{\hskip3pt}lllllllllll} & Loss Function & ERR & P@1 & P@3 & P@5 & nDCG@1 & nDCG@3 & nDCG@5 & AP \\ \hline
\multirow{5}{*}{MQ2007}
& \underline{GBM -- LambdaMART} & 0.3211 & 0.4669 & 0.4397 & 0.4167 & 0.4217 & 0.4243 & 0.4285 & 0.4646 \\
& GBM -- Pointwise KL (Binomial) & \textbf{0.3233}  & 0.4752  & 0.4354  & 0.4167  & 0.4303  & 0.4207  & 0.4275  & 0.4591$\downarrow$  \\
& GBM -- Listwise KL (Gaussian) & 0.3219  & 0.4592  & \textbf{0.4399}  & 0.4164  & 0.4152  & 0.4240  & 0.4267  & 0.4595  \\
& \textbf{NN -- Pairwise KL (Gaussian)} & 0.3218  & \textbf{0.4817}  & 0.4381  & \textbf{0.4201}  & \textbf{0.4350}  & \textbf{0.4249}  & \textbf{0.4318}  & \textbf{0.4665}  \\
& NN -- Listwise KL (Gaussian) & 0.3177  & 0.4657  & 0.4332  & 0.4145  & 0.4173  & 0.4192  & 0.4255  & 0.4634  \\\hline\hline
\multirow{5}{*}{MQ2008}
& \underline{GBM -- LambdaMART} & 0.3045 & 0.4413 & 0.3869 & 0.3446 & 0.3858 & 0.4260 & 0.4664 & 0.4746 \\
& \textbf{GBM -- Pointwise KL (Binomial)} & \textbf{0.3072}  & \textbf{0.4439}  & \textbf{0.3941}  & \textbf{0.3464}  & \textbf{0.3935}  & \textbf{0.4325}  & \textbf{0.4690}  & 0.4771  \\
& GBM -- Listwise KL (Gaussian) & 0.2991  & 0.4362  & 0.3852  & 0.3444  & 0.3801  & 0.4214  & 0.4633  & \textbf{0.4775}  \\
& NN -- Pairwise KL (Gaussian) & 0.3019  & 0.4375  & 0.3852  & 0.3398  & 0.3871  & 0.4222  & 0.4603  & 0.4697  \\
& NN -- Listwise KL (Gaussian) & 0.3008  & 0.4349  & 0.3814  & 0.3457  & 0.3827  & 0.4171  & 0.4630  & 0.4729  \\\hline\hline
\multirow{5}{*}{WEB30K}
& \textbf{\underline{GBM -- LambdaMART}} & \textbf{0.3955} & \textbf{0.7918} & 0.7541 & 0.7288 & \textbf{0.5925} & \textbf{0.5711} & \textbf{0.5670} & 0.6299 \\
& GBM -- Pointwise KL (Binomial) & 0.3550$\downarrow$  & 0.7789$\downarrow$  & 0.7503  & 0.7261  & 0.5435$\downarrow$  & 0.5344$\downarrow$  & 0.5343$\downarrow$  & 0.6326$\uparrow$  \\
& GBM -- Listwise KL (Gaussian) & 0.3861$\downarrow$  & \textbf{0.7918}  & \textbf{0.7565}  & \textbf{0.7323}$\uparrow$  & 0.5825$\downarrow$  & 0.5647$\downarrow$  & 0.5615$\downarrow$  & \textbf{0.6347}$\uparrow$  \\
& NN -- Pairwise KL (Gaussian) & 0.3454$\downarrow$  & 0.7753$\downarrow$  & 0.7423$\downarrow$  & 0.7166$\downarrow$  & 0.5315$\downarrow$  & 0.5209$\downarrow$  & 0.5214$\downarrow$  & 0.5971$\downarrow$  \\
& NN -- Listwise KL (Gaussian) & 0.3523$\downarrow$  & 0.7612$\downarrow$  & 0.7258$\downarrow$  & 0.7016$\downarrow$  & 0.5322$\downarrow$  & 0.5152$\downarrow$  & 0.5141$\downarrow$  & 0.5871$\downarrow$  \\\hline\hline
\multirow{5}{*}{OHSUMED}
& \underline{GBM -- LambdaMART} & 0.4704 & 0.5283 & 0.4874 & 0.4906 & 0.4387 & 0.3980 & 0.4037 & 0.4175 \\
& GBM -- Pointwise KL (Binomial) & 0.5036  & 0.5755  & 0.5220  & 0.5000  & 0.4953  & 0.4474  & 0.4330  & 0.4210  \\
& GBM -- Listwise KL (Gaussian) & 0.5139  & 0.5755  & 0.5314  & \textbf{0.5151}  & 0.5000  & 0.4643$\uparrow$  & \textbf{0.4525}$\uparrow$  & \textbf{0.4243}  \\
& NN -- Pairwise KL (Gaussian) & 0.4520  & 0.5377  & 0.4403  & 0.4000$\downarrow$  & 0.4481  & 0.3707  & 0.3481$\downarrow$  & 0.2903$\downarrow$  \\
& \textbf{NN -- Listwise KL (Gaussian)} & \textbf{0.5248}  & \textbf{0.6038}  & \textbf{0.5692}$\uparrow$  & 0.5057  & \textbf{0.5189}  & \textbf{0.4796}$\uparrow$  & 0.4456  & 0.3861$\downarrow$  \\\hline

\end{tabular}}\caption{Performance of different \ac{LETOR} models (decision tree-based Gradient Boosted Machine (GBM) model or the Neural Model (NM)) trained with the best-performing proposed loss functions  averaged over all topics. $\uparrow$ or $\downarrow$ indicate a statistically significant ($\alpha<0.05$) difference with the LambdaMART model trained on the original relevance judgments. Best performance measures per collection are in bold as the loss function with the most best measures per collection.}
\label{tab:lambdamart_comparison}
\end{table*}
Finally, in Table~\ref{tab:lambdamart_comparison} we report the performance comparison between the best performing systems relying on the proposed neural model or on a decision tree-based \ac{GBM} from the LightGBM library. We also report here the performance of a LambdaMART model trained with the LigthGBM library as a baseline. LambdaMART, and in particular its LightGBM implementation is generally considered a very competitive baseline in many other \ac{LETOR} research works~\cite{bruch2019revisiting, bruch2019alternative, pasumarthi2020permutation, bruch2020stochastic, qin2021neural}. We observe that the proposed probabilistic loss functions are in most of the cases cases able to improve the performance of a decision tree-based \ac{GBM} model, surpassing the one of LambdaMART. We also observe that, on the MQ2007 and OHSUMED experimental collections, the proposed neural model outperforms LambdaMART and all the \ac{GBM}-based models by a sizable margin. 

\begin{table}[h!] \centering \resizebox{\textwidth}{!}{\begin{tabular}{@{\hskip3pt}l@{\hskip3pt}@{\hskip3pt}l@{\hskip3pt}lllllllllll} & Optimization Function & ERR & P@1 & P@3 & P@5 & nDCG@1 & nDCG@3 & nDCG@5 & MAP \\ \hline
\multirow{8}{*}{MQ2007}
& DASALC -- Pointwise KL (Binomial) &  -0.0014 &  -0.0047 &  -0.0032 & \textbf{ +0.0032} &  -0.0062 &  -0.0057 &  -0.0014 &  -0.0011 \\
& DASALC -- Pairwise KL (Binomial) & \textbf{ +0.0016} & \textbf{ +0.0065} & \textbf{ +0.0041} & \textbf{ +0.0067}$\uparrow$ & \textbf{ +0.0059} & \textbf{ +0.0027} & \textbf{ +0.0050} & \textbf{ +0.0060}$\uparrow$ \\
& DASALC -- Pairwise KL (Gaussian) & \textbf{ +0.0007} & \textbf{ +0.0035} & \textbf{ +0.0047} & \textbf{ +0.0080}$\uparrow$ & \textbf{ +0.0027} & \textbf{ +0.0038} & \textbf{ +0.0062} & \textbf{ +0.0062}$\uparrow$ \\
& DASALC -- Listwise KL (Gaussian) &  -0.0015 &  -0.0089 &  -0.0006 & \textbf{ +0.0020} &  -0.0080 &  -0.0014 & \textbf{ +0.0004} &  -0.0002 \\\cdashline{2-10}
& TRANSFORMER -- Pointwise KL (Binomial) & \textbf{ +0.0099}$\uparrow$ & \textbf{ +0.0035} & \textbf{ +0.0106}$\uparrow$ &  -0.0007 & \textbf{ +0.0098} & \textbf{ +0.0112}$\uparrow$ & \textbf{ +0.0033} &  -0.0005 \\
& TRANSFORMER -- Pairwise KL (Binomial) & \textbf{ +0.0084}$\uparrow$ & \textbf{ +0.0047} & \textbf{ +0.0091} & \textbf{ +0.0041} & \textbf{ +0.0095} & \textbf{ +0.0078} & \textbf{ +0.0033} & \textbf{ +0.0025} \\
& TRANSFORMER -- Pairwise KL (Gaussian) &  -0.0022 &  -0.0154 &  -0.0022 &  -0.0045 &  -0.0115 &  -0.0060 &  -0.0087 &  -0.0039 \\
& TRANSFORMER -- Listwise KL (Gaussian) & \textbf{ +0.0015} & \textbf{ +0.0024} & \textbf{ +0.0002} &  -0.0022 & \textbf{ +0.0038} & \textbf{ +0.0007} &  -0.0012 & \textbf{ +0.0001} \\\hline\hline
\multirow{8}{*}{MQ2008}
& DASALC -- Pointwise KL (Binomial) & \textbf{ +0.0116}$\uparrow$ & \textbf{ +0.0255} & \textbf{ +0.0136}$\uparrow$ & \textbf{ +0.0128}$\uparrow$ & \textbf{ +0.0236} & \textbf{ +0.0172}$\uparrow$ & \textbf{ +0.0226}$\uparrow$ & \textbf{ +0.0179}$\uparrow$ \\
& DASALC -- Pairwise KL (Binomial) & \textbf{ +0.0162}$\uparrow$ & \textbf{ +0.0306}$\uparrow$ & \textbf{ +0.0234}$\uparrow$ & \textbf{ +0.0158}$\uparrow$ & \textbf{ +0.0344}$\uparrow$ & \textbf{ +0.0284}$\uparrow$ & \textbf{ +0.0275}$\uparrow$ & \textbf{ +0.0236}$\uparrow$ \\
& DASALC -- Pairwise KL (Gaussian) & \textbf{ +0.0114}$\uparrow$ & \textbf{ +0.0230} & \textbf{ +0.0170}$\uparrow$ & \textbf{ +0.0089}$\uparrow$ & \textbf{ +0.0249}$\uparrow$ & \textbf{ +0.0210}$\uparrow$ & \textbf{ +0.0201}$\uparrow$ & \textbf{ +0.0181}$\uparrow$ \\
& DASALC -- Listwise KL (Gaussian) & \textbf{ +0.0160}$\uparrow$ & \textbf{ +0.0268}$\uparrow$ & \textbf{ +0.0238}$\uparrow$ & \textbf{ +0.0171}$\uparrow$ & \textbf{ +0.0268}$\uparrow$ & \textbf{ +0.0277}$\uparrow$ & \textbf{ +0.0279}$\uparrow$ & \textbf{ +0.0224}$\uparrow$ \\\cdashline{2-10}
& TRANSFORMER -- Pointwise KL (Binomial) & \textbf{ +0.0119}$\uparrow$ & \textbf{ +0.0242} & \textbf{ +0.0162}$\uparrow$ & \textbf{ +0.0056} & \textbf{ +0.0198} & \textbf{ +0.0231}$\uparrow$ & \textbf{ +0.0170}$\uparrow$ & \textbf{ +0.0146}$\uparrow$ \\
& TRANSFORMER -- Pairwise KL (Binomial) & \textbf{ +0.0129}$\uparrow$ & \textbf{ +0.0293}$\uparrow$ & \textbf{ +0.0174}$\uparrow$ & \textbf{ +0.0074} & \textbf{ +0.0281}$\uparrow$ & \textbf{ +0.0235}$\uparrow$ & \textbf{ +0.0178}$\uparrow$ & \textbf{ +0.0143}$\uparrow$ \\
& TRANSFORMER -- Pairwise KL (Gaussian) & \textbf{ +0.0022} & \textbf{ +0.0051} & \textbf{ +0.0230}$\uparrow$ & \textbf{ +0.0018} &  -0.0064 & \textbf{ +0.0173}$\uparrow$ & \textbf{ +0.0051} & \textbf{ +0.0046} \\
& TRANSFORMER -- Listwise KL (Gaussian) & \textbf{ +0.0117}$\uparrow$ & \textbf{ +0.0281}$\uparrow$ & \textbf{ +0.0170}$\uparrow$ & \textbf{ +0.0094}$\uparrow$ & \textbf{ +0.0191} & \textbf{ +0.0239}$\uparrow$ & \textbf{ +0.0213}$\uparrow$ & \textbf{ +0.0190}$\uparrow$ \\\hline\hline
\multirow{8}{*}{WEB30K}
& DASALC -- Pointwise KL (Binomial) &  -0.0083$\downarrow$ &  -0.0262$\downarrow$ &  -0.0152$\downarrow$ &  -0.0144$\downarrow$ &  -0.0192$\downarrow$ &  -0.0175$\downarrow$ &  -0.0173$\downarrow$ &  -0.0023$\downarrow$ \\
& DASALC -- Pairwise KL (Binomial) &  -0.0120$\downarrow$ & \textbf{ +0.0165}$\uparrow$ & \textbf{ +0.0209}$\uparrow$ & \textbf{ +0.0216}$\uparrow$ & \textbf{ +0.0028} & \textbf{ +0.0062}$\uparrow$ & \textbf{ +0.0077}$\uparrow$ & \textbf{ +0.0158}$\uparrow$ \\
& DASALC -- Pairwise KL (Gaussian) &  -0.0169$\downarrow$ & \textbf{ +0.0094}$\uparrow$ & \textbf{ +0.0168}$\uparrow$ & \textbf{ +0.0182}$\uparrow$ &  -0.0074 &  -0.0009 & \textbf{ +0.0020} & \textbf{ +0.0148}$\uparrow$ \\
& DASALC -- Listwise KL (Gaussian) &  -0.0067$\downarrow$ &  -0.0263$\downarrow$ &  -0.0175$\downarrow$ &  -0.0138$\downarrow$ &  -0.0163$\downarrow$ &  -0.0139$\downarrow$ &  -0.0125$\downarrow$ &  -0.0039$\downarrow$ \\\cdashline{2-10}
& TRANSFORMER -- Pointwise KL (Binomial) &  -0.0238$\downarrow$ & \textbf{ +0.0138}$\uparrow$ &  -0.0028 &  -0.0146$\downarrow$ &  -0.0099$\downarrow$ &  -0.0193$\downarrow$ &  -0.0243$\downarrow$ &  -0.0082$\downarrow$ \\
& TRANSFORMER -- Pairwise KL (Binomial) &  -0.0440$\downarrow$ & \textbf{ +0.0135}$\uparrow$ & \textbf{ +0.0142}$\uparrow$ & \textbf{ +0.0101}$\uparrow$ &  -0.0386$\downarrow$ &  -0.0235$\downarrow$ &  -0.0197$\downarrow$ & \textbf{ +0.0042}$\uparrow$ \\
& TRANSFORMER -- Pairwise KL (Gaussian) &  -0.0419$\downarrow$ &  -0.0041 &  -0.0001 &  -0.0003 &  -0.0416$\downarrow$ &  -0.0336$\downarrow$ &  -0.0279$\downarrow$ & \textbf{ +0.0019}$\uparrow$ \\
& TRANSFORMER -- Listwise KL (Gaussian) &  -0.1000$\downarrow$ &  -0.0798$\downarrow$ &  -0.0826$\downarrow$ &  -0.0786$\downarrow$ &  -0.1242$\downarrow$ &  -0.1136$\downarrow$ &  -0.1068$\downarrow$ &  -0.0328$\downarrow$ \\\hline\hline
\multirow{8}{*}{OHSUMED}
& DASALC -- Pointwise KL (Binomial) & \textbf{ +0.0232} &  0.0000 & \textbf{ +0.0220} & \textbf{ +0.0208} & \textbf{ +0.0189} & \textbf{ +0.0350} & \textbf{ +0.0354} & \textbf{ +0.0328}$\uparrow$ \\
& DASALC -- Pairwise KL (Binomial) & \textbf{ +0.0022} & \textbf{ +0.0094} &  -0.0346 &  -0.0547$\downarrow$ & \textbf{ +0.0283} &  -0.0141 &  -0.0260 &  -0.0107 \\
& DASALC -- Pairwise KL (Gaussian) &  -0.0176 &  -0.0189 &  -0.0629 &  -0.0717$\downarrow$ & \textbf{ +0.0047} &  -0.0408 &  -0.0480$\downarrow$ &  -0.0140 \\
& DASALC -- Listwise KL (Gaussian) & \textbf{ +0.0266} & \textbf{ +0.0472} &  -0.0094 &  -0.0189 & \textbf{ +0.0566} & \textbf{ +0.0178} & \textbf{ +0.0050} & \textbf{ +0.0026} \\\cdashline{2-10}
& TRANSFORMER -- Pointwise KL (Binomial) & \textbf{ +0.1563}$\uparrow$ & \textbf{ +0.1321}$\uparrow$ & \textbf{ +0.1478}$\uparrow$ & \textbf{ +0.1434}$\uparrow$ & \textbf{ +0.1745}$\uparrow$ & \textbf{ +0.1604}$\uparrow$ & \textbf{ +0.1573}$\uparrow$ & \textbf{ +0.0825}$\uparrow$ \\
& TRANSFORMER -- Pairwise KL (Binomial) & \textbf{ +0.0530} &  -0.0189 & \textbf{ +0.0409} & \textbf{ +0.0434} & \textbf{ +0.0283} & \textbf{ +0.0477} & \textbf{ +0.0474} & \textbf{ +0.0393}$\uparrow$ \\
& TRANSFORMER -- Pairwise KL (Gaussian) & \textbf{ +0.0845}$\uparrow$ & \textbf{ +0.0755} & \textbf{ +0.0786} & \textbf{ +0.0302} & \textbf{ +0.1038} & \textbf{ +0.0909} & \textbf{ +0.0673} & \textbf{ +0.0298} \\
& TRANSFORMER -- Listwise KL (Gaussian) & \textbf{ +0.1485}$\uparrow$ & \textbf{ +0.1132} & \textbf{ +0.1604}$\uparrow$ & \textbf{ +0.1358}$\uparrow$ & \textbf{ +0.1604}$\uparrow$ & \textbf{ +0.1559}$\uparrow$ & \textbf{ +0.1441}$\uparrow$ & \textbf{ +0.0768}$\uparrow$ \\\hline
\end{tabular}}\vspace{5pt}\caption{{Performance of different \ac{LETOR} models trained with the proposed loss functions. We indicate with $\uparrow$ or $\downarrow$ a statistically significant (p-value $<0.05$) difference with the performance obtained by the same model trained with the ApproxNDCG loss function on the original relevance judgements available in each dataset. We indicate in bold all the cases where we observe a performance improvement over the respective baseline.}}\label{tab:other_b}
\end{table}

{
To conclude our evaluation, in Table~\ref{tab:other_b} we report a comparison of the performance improvements obtained using the proposed loss functions to train two state-of-the-art models, i.e. \ac{DASALC}~\cite{qin2021neural}  and a simpler transformer model similar to the one used in ~\cite{pobrotyn2020context}. More specifically, we report -- for each model -- the performance difference when training it using one of the proposed loss functions or the ApproxNDCG loss. From the results reported in Table~\ref{tab:other_b}, we observe that in the majority of our tests the proposed family of loss functions allows a better performance of both \ac{DASALC} and the transformer model compared to the ApproxNDCG loss, especially on the MQ2008 collection. On the MQ2007 collection, we observe performance improvements on all evaluation measures, with some differences between the two models. In this case, the two best loss functions to train the \ac{DASALC} model are the pairwise ones, while the transformer model benefits more from the pointwise and listwise losses. This effect is observed because of the simpler nature of the latter model and its lower number of parameters to train. On the MSLR-WEB30K collection, we observe fewer performance improvements compared to other datasets. However, the performance differences are more similar across models. Here, the best loss function is the Pairwise KL (Binomial), which leads to statistically significant performance improvements for both the considered neural \ac{LETOR} models. Finally, on the OHSUMED collection we observe a significant performance improvement on almost all performance measures when training the transformer model with any of the proposed loss functions. On the other hand, the \ac{DASALC} model shows fewer performance improvements, likely because of the higher model complexity combined with the small number of topics in this collection. Hence, the best loss functions in this case are the Listwise KL (Gaussian) and the Pointwise KL (Binomial), the same that performed the best on the proposed transformer-based neural \ac{LETOR} model.\\
Observing all the above results, we notice that the performance of different models trained with the proposed loss functions varies according to the experimental collection used. Indeed, we conducted our evaluation selecting a number of datasets with different characteristics to show all the strengths and weaknesses of each of the proposed loss functions and to show the best scenarios where they can be employed. For example, for what concerns the MQ2007 and M2008 collections -- with 1,700 and 800 topics, respectively -- we always ranked 128 (or less) documents for each topic. On the other hand, for the MSLR-WEB30K and OHSUMED collection -- with 30,000 and 106 topic each -- we considered a subset of size 128 (or smaller if fewer documents were available) of all documents provided for each topic in the dataset. We selected these subsets by ranking all the available documents for each topic with a lambdaMART model, and then then discarding the items with a rank higher than 128.\\
As a consequence of the diversity of the considered collections, we observe a few differences in the performance of the proposed loss functions in each of our experiments. On medium-sized collections -- where documents were not filtered prior to ranking them -- such as the MQ2007 and MQ2008, we observe overall a similar performance of all the proposed pointwise, pairwise and listwise loss functions, with sizeable differences noticeable only with certain evaluation metrics such as P@\{1-5\} and nDCG@\{1-5\}. 
On the MSLR-WEB30K dataset we observe a similar trend, here however we notice a more sizeable performance difference between the pairwise and listwise loss functions we propose and the pointwise variant -- especially when using them to train the \ac{DASALC} and the proposed \ac{LETOR} model. In this case, the larger availability of training data and the document filtering step we applied before reranking contribute to the better performance of more complex training strategies -- i.e. pairwise and listwise loss functions.\\
Finally, on the OHSUMED collection -- the smallest of all the datasets we considered -- we observe a different situation. Here, despite the small number of topics available, the best-performing loss function is the Listwise KL (Gaussian). This is likely due to the document pruning step we perform prior to ranking. This step promotes the selection of documents which are more similar to each other than in the previous cases and therefore gives an advantage to loss functions which compare multiple items at a time, i.e. the proposed listwise loss function. The second best performing probabilistic loss function however is the pointwise KL one. Indeed, this loss function allows a \ac{LETOR} model to learn better than other training strategies when a few training examples are available -- since it considers each document-topic pair as a valid training data-point, simplifying the training objective.
}
\section{Conclusions and Future Work} \label{sec:conclusion}
{
We presented different strategies to train a \ac{LETOR} model relying on relevance judgments distributions. We introduced five different loss functions relying on the \ac{KL} divergence between distributions, opening new possibilities for the training of \ac{LETOR} models. The proposed loss functions were evaluated on a newly proposed neural model, two transformer-based neural \ac{LETOR} systems, and on a decision tree-based \ac{GBM} model -- the same model employed by the popular LambdaMART algorithm~\cite{burges2010ranknet} -- over a number of experimental collections of different sizes.
}

We compared the performance of the proposed loss functions to the most representative loss functions in the \ac{IR} domain: the pointwise Mean Squared Error (MSE) loss~\cite{liu2018leveraging}, the pairwise Hinge loss~\cite{guo2016deep}, the listwise ApproxNDCG loss~\cite{qin2010general} with and without the Stochastic Treatment (ST) proposed in~\cite{bruch2020stochastic} and the ListMLE loss~\cite{xia2008listwise}.
In our experiments, the proposed loss functions outperformed the aforementioned baselines in several cases and gave a significant performance boost to \ac{LETOR} approaches -- especially the ones based on neural models -- allowing them to also outperform other strong baselines in the \ac{LETOR} domain such as the LightGBM implementation of LambdaMART~\cite{burges2010ranknet, qin2021neural}.

We also evaluated the option of training a neural \ac{LETOR} model simulating the distribution of relevance judgments for each document-topic pair in the training data.
The results from this experiment further confirmed our hypothesis on the utility of using relevance judgments distributions to train a \ac{LETOR} model, showing performance improvements across different measures.

Finally, we conducted a crowdsourcing experiment on the COVID-19 MLIA collection, building a new \ac{LETOR} collection with real relevance judgments distributions. We share this collection and labels to be used for the development and evaluation of other \ac{LETOR} approaches that will follow the proposed training paradigm. These experiments  consolidated our hypothesis and showed encouraging results on the usage of probabilistic loss functions also on this dataset.
{
As future work, we plan to further develop the proposed neural architecture to take advantage of probability distributions on model weights -- i.e. employing Bayesian neural layers~\cite{korattikara2015bayesian} -- and to evaluate the performance of the proposed loss functions to train a model based on implicit user feedback signals such as clicks and dwell time.}

\acrodef{3G}[3G]{Third Generation Mobile System}
\acrodef{5S}[5S]{Streams, Structures, Spaces, Scenarios, Societies}
\acrodef{AAAI}[AAAI]{Association for the Advancement of Artificial Intelligence}
\acrodef{AAL}[AAL]{Annotation Abstraction Layer}
\acrodef{AAM}[AAM]{Automatic Annotation Manager}
\acrodef{ACLIA}[ACLIA]{Advanced Cross-Lingual Information Access}
\acrodef{ACM}[ACM]{Association for Computing Machinery}
\acrodef{ADSL}[ADSL]{Asymmetric Digital Subscriber Line}
\acrodef{ADUI}[ADUI]{ADministrator User Interface}
\acrodef{AIP}[AIP]{Archival Information Package}
\acrodef{AJAX}[AJAX]{Asynchronous JavaScript Technology and \acs{XML}}
\acrodef{ALU}[ALU]{Aritmetic-Logic Unit}
\acrodef{AMUSID}[AMUSID]{Adaptive MUSeological IDentity-service}
\acrodef{ANOVA}[ANOVA]{ANalysis Of VAriance}
\acrodef{ANSI}[ANSI]{American National Standards Institute}
\acrodef{AP}[AP]{Average Precision}
\acrodef{APC}[APC]{AP Correlation}
\acrodef{API}[API]{Application Program Interface}
\acrodef{AR}[AR]{Address Register}
\acrodef{AS}[AS]{Annotation Service}
\acrodef{ASAP}[ASAP]{Adaptable Software Architecture Performance}
\acrodef{ASI}[ASI]{Annotation Service Integrator}
\acrodef{ASM}[ASM]{Annotation Storing Manager}
\acrodef{ASR}[ASR]{Automatic Speech Recognition}
\acrodef{ASUI}[ASUI]{ASsessor User Interface}
\acrodef{ATIM}[ATIM]{Annotation Textual Indexing Manager}
\acrodef{AUC}[AUC]{Area Under the ROC Curve}
\acrodef{AUI}[AUI]{Administrative User Interface}
\acrodef{AWARE}[AWARE]{Assessor-driven Weighted Averages for Retrieval Evaluation}
\acrodef{BANKS-I}[BANKS-I]{Browsing ANd Keyword Searching I}
\acrodef{BANKS-II}[BANKS-II]{Browsing ANd Keyword Searching II}
\acrodef{BERT}[BERT]{Bidirectional Encoder Representations from Transformers}
\acrodef{bpref}[bpref]{Binary Preference}
\acrodef{BNF}[BNF]{Backus and Naur Form}
\acrodef{BRICKS}[BRICKS]{Building Resources for Integrated Cultural Knowledge Services}
\acrodef{CAN}[CAN]{Content Addressable Netword}
\acrodef{CAS}[CAS]{Content-And-Structure}
\acrodef{CBSD}[CBSD]{Component-Based Software Developlement}
\acrodef{CBSE}[CBSE]{Component-Based Software Engineering}
\acrodef{CB-SPE}[CB-SPE]{Component-Based \acs{SPE}}
\acrodef{CD}[CD]{Collaboration Diagram}
\acrodef{CD}[CD]{Compact Disk}
\acrodef{CDF}[CDF]{Cumulative Distribution Function}
\acrodef{CENL}[CENL]{Conference of European National Librarians}
\acrodef{CIDOC CRM}[CIDOC CRM]{CIDOC Conceptual Reference Model}
\acrodef{CIR}[CIR]{Current Instruction Register}
\acrodef{CIRCO}[CIRCO]{Coordinated Information Retrieval Components Orchestration}
\acrodef{CG}[CG]{Cumulated Gain}
\acrodef{CLAIRE}[CLAIRE]{Combinatorial visuaL Analytics system for Information Retrieval Evaluation}
\acrodef{CLEF}[CLEF]{Conference and Labs of the Evaluation Forum}
\acrodef{CLIR}[CLIR]{Cross Language Information Retrieval}
\acrodef{CMS}[CMS]{Content Management System}
\acrodef{CMT}[CMT]{Campaign Management Tool}
\acrodef{CNR}[CNR]{Italian National Council of Research}
\acrodef{CO}[CO]{Content-Only}
\acrodef{COD}[COD]{Code On Demand}
\acrodef{CODATA}[CODATA]{Committee on Data for Science and Technology}
\acrodef{COLLATE}[COLLATE]{Collaboratory for Annotation Indexing and Retrieval of Digitized Historical Archive Material}
\acrodef{CP}[CP]{Characteristic Pattern}
\acrodef{CPE}[CPE]{Control Processor Element}
\acrodef{CPU}[CPU]{Central Processing Unit}
\acrodef{CQL}[CQL]{Contextual Query Language}
\acrodef{CRP}[CRP]{Cumulated Relative Position}
\acrodef{CRUD}[CRUD]{Create--Read--Update--Delete}
\acrodef{CS}[CS]{Characteristic Structure}
\acrodef{CSM}[CSM]{Campaign Storing Manager}
\acrodef{CSS}[CSS]{Cascading Style Sheets}
\acrodef{CU}[CU]{Control Unit}
\acrodef{CUI}[CUI]{Client User Interface}
\acrodef{CV}[CV]{Cross-Validation}
\acrodef{DAFFODIL}[DAFFODIL]{Distributed Agents for User-Friendly Access of Digital Libraries}
\acrodef{DAO}[DAO]{Data Access Object}
\acrodef{DARE}[DARE]{Drawing Adequate REpresentations}
\acrodef{DARPA}[DARPA]{Defense Advanced Research Projects Agency}
\acrodef{DAS}[DAS]{Distributed Annotation System}
\acrodef{DASALC}[DASALC]{Data Augmented Self-Attentive Latent Cross model}
\acrodef{DB}[DB]{DataBase}
\acrodef{DBMS}[DBMS]{DataBase Management System}
\acrodef{DC}[DC]{Dublin Core}
\acrodef{DCG}[DCG]{Discounted Cumulated Gain}
\acrodef{DCMI}[DCMI]{Dublin Core Metadata Initiative}
\acrodef{DCV}[DCV]{Document Cut--off Value}
\acrodef{DD}[DD]{Deployment Diagram}
\acrodef{DDC}[DDC]{Dewey Decimal Classification}
\acrodef{DDS}[DDS]{Direct Data Structure}
\acrodef{DF}[DF]{Degrees of Freedom}
\acrodef{DFR}[DFR]{Divergence From Randomness}
\acrodef{DHT}[DHT]{Distributed Hash Table}
\acrodef{DI}[DI]{Digital Image}
\acrodef{DIKW}[DIKW]{Data, Information, Knowledge, Wisdom}
\acrodef{DIL}[DIL]{\acs{DIRECT} Integration Layer}
\acrodef{DiLAS}[DiLAS]{Digital Library Annotation Service}
\acrodef{DIRECT}[DIRECT]{Distributed Information Retrieval Evaluation Campaign Tool}
\acrodef{DKMS}[DKMS]{Data and Knowledge Management System}
\acrodef{DL}[DL]{Digital Library}
\acrodefplural{DL}[DL]{Digital Libraries}
\acrodef{DLMS}[DLMS]{Digital Library Management System}
\acrodef{DLOG}[DL]{Description Logics}
\acrodef{DLS}[DLS]{Digital Library System}
\acrodef{DLSS}[DLSS]{Digital Library Service System}
\acrodef{DO}[DO]{Digital Object}
\acrodef{DOI}[DOI]{Digital Object Identifier}
\acrodef{DOM}[DOM]{Document Object Model}
\acrodef{DoMDL}[DoMDL]{Document Model for Digital Libraries}
\acrodef{DPBF}[DPBF]{Dynamic Programming Best-First}
\acrodef{DR}[DR]{Data Register}
\acrodef{DRMM}[DRMM]{Deep Relevance Matching Model}
\acrodef{DRIVER}[DRIVER]{Digital Repository Infrastructure Vision for European Research}
\acrodef{DTD}[DTD]{Document Type Definition}
\acrodef{DVD}[DVD]{Digital Versatile Disk}
\acrodef{EAC-CPF}[EAC-CPF]{Encoded Archival Context for Corporate Bodies, Persons, and Families}
\acrodef{EAD}[EAD]{Encoded Archival Description}
\acrodef{EAN}[EAN]{International Article Number}
\acrodef{ECD}[ECD]{Enhanced Contenty Delivery}
\acrodef{ECDL}[ECDL]{European Conference on Research and Advanced Technology for Digital Libraries}
\acrodef{ECIR}[ECIR]{European Conference on Information Retrieval}
\acrodef{EDM}[EDM]{Europeana Data Model}
\acrodef{EG}[EG]{Execution Graph}
\acrodef{ELDA}[ELDA]{Evaluation and Language resources Distribution Agency}
\acrodef{ELRA}[ELRA]{European Language Resources Association}
\acrodef{EM}[EM]{Expectation Maximization}
\acrodef{EMMA}[EMMA]{Extensible MultiModal Annotation}
\acrodef{EPROM}[EPROM]{Erasable Programmable \acs{ROM}}
\acrodef{EQNM}[EQNM]{Extended Queueing Network Model}
\acrodef{ER}[ER]{Entity--Relationship}
\acrodef{ERR}[ERR]{Expected Reciprocal Rank}
\acrodef{ETL}[ETL]{Extract-Transform-Load}
\acrodef{FAST}[FAST]{Flexible Annotation Service Tool}
\acrodef{FFNN}[FFNN]{Feed-Forward Neural Network}
\acrodef{FIFO}[FIFO]{First-In / First-Out}
\acrodef{FIRE}[FIRE]{Forum for Information Retrieval Evaluation}
\acrodef{FN}[FN]{False Negative}
\acrodef{FNR}[FNR]{False Negative Rate}
\acrodef{FOAF}[FOAF]{Friend of a Friend}
\acrodef{FORESEE}[FORESEE]{FOod REcommentation sErvER}
\acrodef{FP}[FP]{False Positive}
\acrodef{FPR}[FPR]{False Positive Rate}
\acrodef{GIF}[GIF]{Graphics Interchange Format}
\acrodef{GIR}[GIR]{Geografic Information Retrieval}
\acrodef{GAP}[GAP]{Graded Average Precision}
\acrodef{GBM}[GBM]{Gradient Boosting Machine}
\acrodef{GLM}[GLM]{General Linear Model}
\acrodef{GLMM}[GLMM]{General Linear Mixed Model}
\acrodef{GMAP}[GMAP]{Geometric Mean Average Precision}
\acrodef{GoP}[GoP]{Grid of Points}
\acrodef{GPRS}[GPRS]{General Packet Radio Service}
\acrodef{gRBP}[gRBP]{Graded Rank-Biased Precision}
\acrodef{GTIN}[GTIN]{Global Trade Item Number}
\acrodef{GUI}[GUI]{Graphical User Interface}
\acrodef{GW}[GW]{Gateway}
\acrodef{HCI}[HCI]{Human Computer Interaction}
\acrodef{HDS}[HDS]{Hybrid Data Structure}
\acrodef{HIR}[HIR]{Hypertext Information Retrieval}
\acrodef{HIT}[HIT]{Human Intelligent Task}
\acrodef{HITS}[HITS]{Hyperlink-Induced Topic Search}
\acrodef{HTML}[HTML]{HyperText Markup Language}
\acrodef{HTTP}[HTTP]{HyperText Transfer Protocol}
\acrodef{HSD}[HSD]{Honestly Significant Difference}
\acrodef{ICA}[ICA]{International Council on Archives}
\acrodef{ICSU}[ICSU]{International Council for Science}
\acrodef{IDF}[IDF]{Inverse Document Frequency}
\acrodef{IDS}[IDS]{Inverse Data Structure}
\acrodef{IEEE}[IEEE]{Institute of Electrical and Electronics Engineers}
\acrodef{IEI}[IEI]{Istituto della Enciclopedia Italiana fondata da Giovanni Treccani}
\acrodef{IETF}[IETF]{Internet Engineering Task Force}
\acrodef{IMS}[IMS]{Information Management System}
\acrodef{IMSPD}[IMS]{Information Management Systems Research Group}
\acrodef{indAP}[indAP]{Induced Average Precision}
\acrodef{infAP}[infAP]{Inferred Average Precision}
\acrodef{INEX}[INEX]{INitiative for the Evaluation of \acs{XML} Retrieval}
\acrodef{INS-M}[INS-M]{Inverse Set Data Model}
\acrodef{INTR}[INTR]{Interrupt Register}
\acrodef{IP}[IP]{Internet Protocol}
\acrodef{IPSA}[IPSA]{Imaginum Patavinae Scientiae Archivum}
\acrodef{IR}[IR]{Information Retrieval}
\acrodef{IRON}[IRON]{Information Retrieval ON}
\acrodef{IRON2}[IRON$^2$]{Information Retrieval On aNNotations}
\acrodef{IRON-SAT}[IRON-SAT]{\acs{IRON} - Statistical Analysis Tool}
\acrodef{IRS}[IRS]{Information Retrieval System}
\acrodef{ISAD(G)}[ISAD(G)]{International Standard for Archival Description (General)}
\acrodef{ISBN}[ISBN]{International Standard Book Number}
\acrodef{ISIS}[ISIS]{Interactive SImilarity Search}
\acrodef{ISJ}[ISJ]{Interactive Searching and Judging}
\acrodef{ISO}[ISO]{International Organization for Standardization}
\acrodef{ITU}[ITU]{International Telecommunication Union }
\acrodef{ITU-T}[ITU-T]{Telecommunication Standardization Sector of \acs{ITU}}
\acrodef{IV}[IV]{Information Visualization}
\acrodef{JAN}[JAN]{Japanese Article Number}
\acrodef{JDBC}[JDBC]{Java DataBase Connectivity}
\acrodef{JMB}[JMB]{Java--Matlab Bridge}
\acrodef{JPEG}[JPEG]{Joint Photographic Experts Group}
\acrodef{JSON}[JSON]{JavaScript Object Notation}
\acrodef{JSP}[JSP]{Java Server Pages}
\acrodef{JTE}[JTE]{Java-Treceval Engine}
\acrodef{KDE}[KDE]{Kernel Density Estimation}
\acrodef{KL}[KL]{Kullback–Leibler}
\acrodef{KLD}[KLD]{Kullback-Leibler Divergence}
\acrodef{KLAPER}[KLAPER]{Kernel LAnguage for PErformance and Reliability analysis}
\acrodef{LAM}[LAM]{Libraries, Archives, and Museums}
\acrodef{LAM2}[LAM]{Logistic Average Misclassification}
\acrodef{LAN}[LAN]{Local Area Network}
\acrodef{LD}[LD]{Linked Data}
\acrodef{LEAF}[LEAF]{Linking and Exploring Authority Files}
\acrodef{LETOR}[LETOR]{LEarning TO Rank}
\acrodef{LIDO}[LIDO]{Lightweight Information Describing Objects}
\acrodef{LIFO}[LIFO]{Last-In / First-Out}
\acrodef{LM}[LM]{Language Model}
\acrodef{LMT}[LMT]{Log Management Tool}
\acrodef{LOD}[LOD]{Linked Open Data}
\acrodef{LODE}[LODE]{Linking Open Descriptions of Events}
\acrodef{LpO}[LpO]{Leave-$p$-Out}
\acrodef{LRM}[LRM]{Local Relational Model}
\acrodef{LRU}[LRU]{Last Recently Used}
\acrodef{LS}[LS]{Lexical Signature}
\acrodef{LSM}[LSM]{Log Storing Manager}
\acrodef{LUG}[LUG]{Lexical Unit Generator}
\acrodef{MA}[MA]{Mobile Agent}
\acrodef{MA}[MA]{Moving Average}
\acrodef{MACS}[MACS]{Multilingual ACcess to Subjects}
\acrodef{MADCOW}[MADCOW]{Multimedia Annotation of Digital Content Over the Web}
\acrodef{MAD}[MAD]{Mean Assessed Documents}
\acrodef{MADP}[MADP]{Mean Assessed Documents Precision}
\acrodef{MADS}[MADS]{Metadata Authority Description Standard}
\acrodef{MAP}[MAP]{Mean Average Precision}
\acrodef{MARC}[MARC]{Machine Readable Cataloging}
\acrodef{MATTERS}[MATTERS]{MATlab Toolkit for Evaluation of information Retrieval Systems}
\acrodef{MDA}[MDA]{Model Driven Architecture}
\acrodef{MDD}[MDD]{Model-Driven Development}
\acrodef{METS}[METS]{Metadata Encoding and Transmission Standard}
\acrodef{MIDI}[MIDI]{Musical Instrument Digital Interface}
\acrodef{MIME}[MIME]{Multipurpose Internet Mail Extensions}
\acrodef{ML}[ML]{Machine Learning}
\acrodef{MLIA}[MLIA]{MultiLingual Information Access}
\acrodef{MM}[MM]{Machinery Model}
\acrodef{MMU}[MMU]{Memory Management Unit}
\acrodef{MODS}[MODS]{Metadata Object Description Schema}
\acrodef{MOF}[MOF]{Meta-Object Facility}
\acrodef{MP}[MP]{Markov Precision}
\acrodef{MPEG}[MPEG]{Motion Picture Experts Group}
\acrodef{MRD}[MRD]{Machine Readable Dictionary}
\acrodef{MRF}[MRF]{Markov Random Field}
\acrodef{MS}[MS]{Mean Squares}
\acrodef{MSAC}[MSAC]{Multilingual Subject Access to Catalogues}
\acrodef{MSE}[MSE]{Mean Squared Error}
\acrodef{MT}[MT]{Machine Translation}
\acrodef{MV}[MV]{Majority Vote}
\acrodef{MVC}[MVC]{Model-View-Controller}
\acrodef{NACSIS}[NACSIS]{NAtional Center for Science Information Systems}
\acrodef{NAP}[NAP]{Network processors Applications Profile}
\acrodef{NCP}[NCP]{Normalized Cumulative Precision}
\acrodef{nCG}[nCG]{Normalized Cumulated Gain}
\acrodef{nCRP}[nCRP]{Normalized Cumulated Relative Position}
\acrodef{nDCG}[nDCG]{normalized Discounted Cumulated Gain}
\acrodef{NESTOR}[NESTOR]{NEsted SeTs for Object hieRarchies}
\acrodef{NIR}[NIR]{Neural Information Retrieval}
\acrodef{NEXI}[NEXI]{Narrowed Extended XPath I}
\acrodef{NII}[NII]{National Institute of Informatics}
\acrodef{NISO}[NISO]{National Information Standards Organization}
\acrodef{NIST}[NIST]{National Institute of Standards and Technology}
\acrodef{NLP}[NLP]{Natural Language Processing}
\acrodef{NP}[NP]{Network Processor}
\acrodef{NR}[NR]{Normalized Recall}
\acrodef{NS-M}[NS-M]{Nested Set Model}
\acrodef{NTCIR}[NTCIR]{NII Testbeds and Community for Information access Research}
\acrodef{NVSM}[NVSM]{Neural Vector Space Model}
\acrodef{OAI}[OAI]{Open Archives Initiative}
\acrodef{OAI-ORE}[OAI-ORE]{Open Archives Initiative Object Reuse and Exchange}
\acrodef{OAI-PMH}[OAI-PMH]{Open Archives Initiative Protocol for Metadata Harvesting}
\acrodef{OAIS}[OAIS]{Open Archival Information System}
\acrodef{OC}[OC]{Operation Code}
\acrodef{OCLC}[OCLC]{Online Computer Library Center}
\acrodef{OMG}[OMG]{Object Management Group}
\acrodef{OLAP}[OLAP]{On-Line Analytical Processing}
\acrodef{OO}[OO]{Object Oriented}
\acrodef{OODB}[OODB]{Object-Oriented \acs{DB}}
\acrodef{OODBMS}[OODBMS]{Object-Oriented \acs{DBMS}}
\acrodef{OPAC}[OPAC]{Online Public Access Catalog}
\acrodef{OQL}[OQL]{Object Query Language}
\acrodef{ORP}[ORP]{Open Relevance Project}
\acrodef{OSIRIS}[OSIRIS]{Open Service Infrastructure for Reliable and Integrated process Support}
\acrodef{P2P}[P2P]{Peer-To-Peer}
\acrodef{PA}[PA]{Performance Analysis}
\acrodef{PAMT}[PAMT]{Pool-Assessment Management Tool}
\acrodef{PASM}[PASM]{Pool-Assessment Storing Manager}
\acrodef{PC}[PC]{Program Counter}
\acrodef{PCP}[PCP]{Pre-Commercial Procurement}
\acrodef{PCR}[PCR]{Peripherical Command Register}
\acrodef{PDA}[PDA]{Personal Digital Assistant}
\acrodef{PDF}[PDF]{Probability Density Function}
\acrodef{PDR}[PDR]{Peripherical Data Register}
\acrodef{POI}[POI]{\acs{PURL}-based Object Identifier}
\acrodef{PoS}[PoS]{Part of Speech}
\acrodef{PPE}[PPE]{Programmable Processing Engine}
\acrodef{PREFORMA}[PREFORMA]{PREservation FORMAts for culture information/e-archives}
\acrodef{PRIMAmob-UML}[PRIMAmob-UML]{mobile \acs{PRIMA-UML}}
\acrodef{PRIMA-UML}[PRIMA-UML]{PeRformance IncreMental vAlidation in \acs{UML}}
\acrodef{PROM}[PROM]{Programmable \acs{ROM}}
\acrodef{PROMISE}[PROMISE]{Participative Research labOratory  for Multimedia and Multilingual Information Systems Evaluation}
\acrodef{pSQL}[pSQL]{propagate \acs{SQL}}
\acrodef{PUI}[PUI]{Participant User Interface}
\acrodef{PURL}[PURL]{Persistent \acs{URL}}
\acrodef{QA}[QA]{Question Answering}
\acrodef{QoS-UML}[QoS-UML]{\acs{UML} Profile for QoS and Fault Tolerance}
\acrodef{RAM}[RAM]{Random Access Memory}
\acrodef{RAMM}[RAM]{Random Access Machine}
\acrodef{RBO}[RBO]{Rank-Biased Overlap}
\acrodef{RBP}[RBP]{Rank-Biased Precision}
\acrodef{RDBMS}[RDBMS]{Relational \acs{DBMS}}
\acrodef{RDF}[RDF]{Resource Description Framework}
\acrodef{ReLU}[ReLU]{Rectified Linear Unit}
\acrodef{REST}[REST]{REpresentational State Transfer}
\acrodef{REV}[REV]{Remote Evaluation}
\acrodef{RFC}[RFC]{Request for Comments}
\acrodef{RIA}[RIA]{Reliable Information Access}
\acrodef{RMSE}[RMSE]{Root Mean Squared Error}
\acrodef{RMT}[RMT]{Run Management Tool}
\acrodef{ROM}[ROM]{Read Only Memory}
\acrodef{ROMIP}[ROMIP]{Russian Information Retrieval Evaluation Seminar}
\acrodef{RoMP}[RoMP]{Rankings of Measure Pairs}
\acrodef{RoS}[RoS]{Rankings of Systems}
\acrodef{RP}[RP]{Relative Position}
\acrodef{RR}[RR]{Reciprocal Rank}
\acrodef{RSM}[RSM]{Run Storing Manager}
\acrodef{RST}[RST]{Rhetorical Structure Theory}
\acrodef{RT-UML}[RT-UML]{\acs{UML} Profile for Schedulability, Performance and Time}
\acrodef{SA}[SA]{Software Architecture}
\acrodef{SAL}[SAL]{Storing Abstraction Layer}
\acrodef{SAMT}[SAMT]{Statistical Analysis Management Tool}
\acrodef{SAN}[SAN]{Sistema Archivistico Nazionale}
\acrodef{SASM}[SASM]{Statistical Analysis Storing Manager}
\acrodef{SD}[SD]{Sequence Diagram}
\acrodef{SE}[SE]{Search Engine}
\acrodef{SEBD}[SEBD]{Convegno Nazionale su Sistemi Evoluti per Basi di Dati}
\acrodef{SFT}[SFT]{Satisfaction--Frustration--Total}
\acrodef{SIGIR}[SIGIR]{ACM SIGIR Conference on Research \& Development in Information Retrieval}
\acrodef{SIL}[SIL]{Service Integration Layer}
\acrodef{SIP}[SIP]{Submission Information Package}
\acrodef{SKOS}[SKOS]{Simple Knowledge Organization System}
\acrodef{SM}[SM]{Software Model}
\acrodef{SMART}[SMART]{System for the Mechanical Analysis and Retrieval of Text}
\acrodef{SoA}[SoA]{Service-oriented Architectures}
\acrodef{SOA}[SOA]{Strength of Association}
\acrodef{SOAP}[SOAP]{Simple Object Access Protocol}
\acrodef{SOM}[SOM]{Self-Organizing Map}
\acrodef{SOTA}[SOTA]{State-Of-The-Art}
\acrodef{SPE}[SPE]{Software Performance Engineering}
\acrodef{SPINA}[SPINA]{Superimposed Peer Infrastructure for iNformation Access}
\acrodef{SPLIT}[SPLIT]{Stemming Program for Language Independent Tasks}
\acrodef{SPOOL}[SPOOL]{Simultaneous Peripheral Operations On Line}
\acrodef{SQL}[SQL]{Structured Query Language}
\acrodef{SR}[SR]{Sliding Ratio}
\acrodef{SR}[SR]{Status Register}
\acrodef{SRU}[SRU]{Search/Retrieve via \acs{URL}}
\acrodef{SS}[SS]{Sum of Squares}
\acrodef{SSTF}[SSTF]{Shortest Seek Time First}
\acrodef{ST}[ST]{Stochastic Treatment}
\acrodef{STAR}[STAR]{Steiner-Tree Approximation in Relationship graphs}
\acrodef{STON}[STON]{STemming ON}
\acrodef{TAC}[TAC]{Text Analysis Conference}
\acrodef{TBG}[TBG]{Time-Biased Gain}
\acrodef{TCP}[TCP]{Transmission Control Protocol}
\acrodef{TEL}[TEL]{The European Library}
\acrodef{TERRIER}[TERRIER]{TERabyte RetrIEveR}
\acrodef{TF}[TF]{Term Frequency}
\acrodef{TFR}[TFR]{True False Rate}
\acrodef{TN}[TN]{True Negative}
\acrodef{TO}[TO]{Transfer Object}
\acrodef{TP}[TP]{True Positve}
\acrodef{TPR}[TPR]{True Positive Rate}
\acrodef{TRAT}[TRAT]{Text Relevance Assessing Task}
\acrodef{TREC}[TREC]{Text REtrieval Conference}
\acrodef{TRECVID}[TRECVID]{TREC Video Retrieval Evaluation}
\acrodef{TTL}[TTL]{Time-To-Live}
\acrodef{UCD}[UCD]{Use Case Diagram}
\acrodef{UDC}[UDC]{Universal Decimal Classification}
\acrodef{uGAP}[uGAP]{User-oriented Graded Average Precision}
\acrodef{UI}[UI]{User Interface}
\acrodef{UML}[UML]{Unified Modeling Language}
\acrodef{UMT}[UMT]{User Management Tool}
\acrodef{UMTS}[UMTS]{Universal Mobile Telecommunication System}
\acrodef{UoM}[UoM]{Utility-oriented Measurement}
\acrodef{UPC}[UPC]{Universal Product Code}
\acrodef{URI}[URI]{Uniform Resource Identifier}
\acrodef{URL}[URL]{Uniform Resource Locator}
\acrodef{URN}[URN]{Uniform Resource Name}
\acrodef{USM}[USM]{User Storing Manager}
\acrodef{VA}[VA]{Visual Analytics}
\acrodef{VATE}[VATE$^2$]{Visual Analytics Tool for Experimental Evaluation}
\acrodef{VIRTUE}[VIRTUE]{Visual Information Retrieval Tool for Upfront Evaluation}
\acrodef{VD}[VD]{Virtual Document}
\acrodef{VIAF}[VIAF]{Virtual International Authority File}
\acrodef{VL}[VL]{Visual Language}
\acrodef{VoIP}[VoIP]{Voice over IP}
\acrodef{VS}[VS]{Visual Sentence}
\acrodef{W3C}[W3C]{World Wide Web Consortium}
\acrodef{WAN}[WAN]{Wide Area Network}
\acrodef{WHO}[WHO]{World Health Organization}
\acrodef{WLAN}[WLAN]{Wireless \acs{LAN}}
\acrodef{WP}[WP]{Work Package}
\acrodef{WS}[WS]{Web Services}
\acrodef{WSD}[WSD]{Word Sense Disambiguation}
\acrodef{WSDL}[WSDL]{Web Services Description Language}
\acrodef{WWW}[WWW]{World Wide Web}
\acrodef{XMI}[XMI]{\acs{XML} Metadata Interchange}
\acrodef{XML}[XML]{eXtensible Markup Language}
\acrodef{XPath}[XPath]{XML Path Language}
\acrodef{XSL}[XSL]{eXtensible Stylesheet Language}
\acrodef{XSL-FO}[XSL-FO]{\acs{XSL} Formatting Objects}
\acrodef{XSLT}[XSLT]{\acs{XSL} Transformations}
\acrodef{YAGO}[YAGO]{Yet Another Great Ontology}
\acrodef{YASS}[YASS]{Yet Another Suffix Stripper}

\bibliographystyle{apalike}
\bibliography{refs}

\end{document}